\documentclass[a4paper,11pt]{article}
\pdfoutput=1 
\usepackage{jcappub} 
\usepackage[T1]{fontenc} 
\usepackage{csquotes}
\usepackage[english]{babel}
\usepackage{multirow}
\usepackage[normalem]{ulem}
\usepackage{url}
\usepackage[utf8]{inputenc}
\usepackage{bm}        
\usepackage{amssymb}   
\usepackage{amsmath}
\usepackage{subfig}
\usepackage{caption,booktabs}
\usepackage{hyperref}

\usepackage{xcolor}
\usepackage{lineno}
\usepackage[symbol]{footmisc}
\usepackage{graphicx} 
\usepackage{bigints}
\usepackage{multirow}

\newcommand{\gal}{{\rm g}}

\newcommand{\de}{\mathrm{d}}

\newcommand{\lum}{\mathcal{L}}

\newcommand{\snr}{{\rm SNR}}

\newcommand{\g}{$\gamma$}
\newcommand{\bi}{\begin{itemize}}
\newcommand{\ei}{\end{itemize}}
\newcommand{\be}{\begin{equation}}
\newcommand{\ee}{\end{equation}}
\newcommand{\nn}{\mathcal{N}}
\newcommand{\Fermi}{\textit{Fermi}-LAT}

\newcommand{\sv}{\langle\sigma_{\rm ann} v\rangle}
\newcommand{\svth}{\langle\sigma_{\rm ann} v\rangle_{\rm th}}
\newcommand{\mdm}{m_{\rm DM}}

\let\Delta\varDelta
\let\Theta\varTheta
\let\Xi\varXi
\let\Pi\varPi
\let\Sigma\varSigma
\let\Upsilon\varUpsilon
\let\Phi\varPhi
\let\Psi\varPsi
\let\Omega\varOmega

\title{Wiener filtering and multi-tracer techniques for dark matter cross-correlations between gamma-ray emission and galaxy catalogs}

\author[1,2,3]{Andrea Rubiola,\footnote[2]{andreamaria.rubiola@unito.it,andreamaria.rubiola@unitn.it}}
\author[2,3,4,5]{Stefano Camera,\footnote[3]{stefano.camera@unito.it}}

\author[2,3]{Nicolao Fornengo\footnote[4]{nicolao.fornengo@unito.it}}

\affiliation[a]{Dipartimento di Fisica, Università degli Studi di Trento, via Sommarive, 14, 38123 Trento, Italy}
 \affiliation[b]{Dipartimento di Fisica, Università degli Studi di Torino, via P. Giuria 1, 10125 Torino, Italy}
\affiliation[c]{INFN – Istituto Nazionale di Fisica Nucleare, Sezione di Torino, via P. Giuria 1, 10125 Torino, Italy}

\affiliation[4]{INAF -- Istituto Nazionale di Astrofisica, Strada Osservatorio 20, 10025 Pino Torinese, Italy}
\affiliation[5]{Department of Physics \& Astronomy, University of the Western Cape, Cape Town 7535, South Africa}

\abstract{Cross-correlations between a gravitational tracer of dark matter and the contribution to the unresolved gamma-ray background (UGRB) from the radiation produced by the annihilation of the particles responsible for the dark matter, have been established as a powerful tool to investigate the particle physics nature of dark matter. Cross-correlations of the UGRB with galaxy catalogs, cluster catalogs and weak lensing have indeed been measured. In this paper we study statistical techniques that could improve the sensitivity of the cross-correlation techniques on the bounds that can be set to the particle dark matter physical properties. The two methods that we investigate are the application of a Wiener filter and the exploitation of the full multi-tracer information. After identifying the optimal strategies, we show that the adoption of a Wiener filter in the cross-correlation analysis can improve the sensitivity to the dark matter annihilation rate by a factor ranging from 2 to 2.5 as compared to the standard analysis where no filter is applied. The inclusion of the full multi-tracer information can improve the sensitivity up to a factor of 5 for dark matter masses below about 50 GeV, the Wiener filter remaining the best option for heavier dark matter.}

\begin{document}
\maketitle
\flushbottom

\section{Introduction}

Gamma-ray emission from dark matter (DM) annihilation or decay in cosmic structures represents one of the most relevant techniques to investigate the nature of dark matter as a new, yet undiscovered, elementary particle. Gamma-rays are one of the main production channels for dark matter composed by weakly interacting massive particles (WIMP), whose mass and interaction strength make them an especially relevant candidate, since their weak scale interactions and mass allow them to be thermally produced in the early Universe in the right amount to explain the current abundance of DM in the Universe (for a recent review, see Ref. \cite{Cirelli:2024ssz}). Gamma-rays, being neutral, directly traces the origin of their production, thus allowing us to correlate the gamma-ray radiation field to the matter distribution in the Universe. This idea, originally proposed in Refs. \cite{camera2013novel,camera2015tomographic,fornengo2014}, relies on the statistical cross-correlation between the gamma-ray cosmic background radiation field with a gravitational tracer of dark matter in the Universe, which can be represented by galaxy or galaxy cluster catalogs, weak gravitational lensing (cosmic shear), CMB lensing \cite{fornengo2015evidence}, neutral hydrogen \cite{pinetti2019}. Even cosmic voids \cite{Arcari:2022zul} have been shown to be relevant, in the case of decaying dark matter.

Astrophysical processes are also expected to produce gamma-rays, thus representing an irreducible background for the dark-matter gamma-ray signal, and here is where the cross-correlation technique becomes relevant. The dominant astrophysical sources that contribute to the {\it cosmological} gamma-ray radiation field are blazars, misaligned active galactic nuclei, flat spectrum radio quasars and star forming galaxies, with pulsars expected to provide a sub-leading contribution. The brightest of these sources have been identified and catalogued, most notably and recently by the Fermi Large Area Telescope (\Fermi). Once these point-like sources are removed from the gamma-ray sky map (and the galactic emission is either masked or subtracted), what remains is the so-called unresolved gamma-ray background (UGRB), which contains photons produced by {\it unresolved} astrophysical sources and, possibly, by dark matter. The UGRB, being produced by gamma-ray sources that are present in the same cosmological structure that form the scaffolding of the Universe, need to have a pattern of fluctuations which is correlated to the pattern of fluctuation of dark matter. The statistical cross-correlation technique leverages on this property and provides a handle in the attempt of separating the DM gamma-ray signal from the emission from astrophysical sources. The cross-correlation technique, in fact, incorporates directly information coming from gravitational tracers of the matter distribution in the Universe, and can exploit differences that arise in angular scale, in the energy spectrum and in the redshift evolution between the cross-correlated fields and among the different classes of gamma-ray emitters \cite{camera2013novel,camera2015tomographic,fornengo2014}. In fact, extragalactic gamma-ray sources typically are observed as point-like entities at gamma-ray energies (due to their compact size as well as the angular resolution of the detector), whereas DM is expected to produce a diffuse signal that traces the large-scale structure of the Universe where dark matter is present. Different classes of astrophysical sources have different spectral energy distributions, which are generically distinct from the energy dependence of the DM-induced emission. Also their respective redshift distributions are at variance: DM gamma-ray emission peaks at low redshift, while the unresolved emission from astrophysical sources has a much broader kernel with respect to cosmological distances {\cite{camera2013novel,camera2015tomographic,fornengo2014}}.

Cross-correlations studies between the UGRB and gravitational tracers of dark matter have been done by using {\it Fermi}-LAT data correlated with several tracers of the large-scale structure: 
weak gravitational lensing  \cite{shirasaki2014cross,shirasaki2016cosmological,troster2017cross,ammazzalorso2020,Thakore_DES_2025}, 
galaxies \cite{xia2011,xia2015tomography,regis2015particle,cuoco2015dark,shirasaki2015cross,cuoco2017tomographic,Ammazzalorso:2018evf,paopiamsap2024constraints}, 
galaxy clusters \cite{branchini2017cross,shirasaki2018correlation,hashimoto2019measurement,colavincenzo2020searching,tan2020bounds} 
and the lensing effect of the cosmic microwave background \cite{fornengo2015evidence}, which traces 
the large-scale distribution of matter across cosmological distances. 
These studies have found a statistically significant evidence for the existence of cross-correlations: with galaxies at the level of $3.5\sigma$ \cite{xia2015tomography,regis2015particle,cuoco2015dark} up to $8-10\sigma$ \cite{paopiamsap2024constraints}, with galaxy clusters at $4.7\sigma$ \cite{branchini2017cross}, with the lensing of the cosmic microwave background at $3.2\sigma$ \cite{fornengo2015evidence} and with the weak gravitational lensing at $5.3\sigma$ \cite{ammazzalorso2020} and at signal-to-noise ratio of 8.9  \cite{Thakore_DES_2025}.
Additional studies of cross-correlations have been presented in Refs. \cite{Ando_2MRS_2014,fornasa2016angular,feng2017planck,Tan:2020fbc,Zhou:2024cld,Pinetti:2025qpi}, including the proposal to use the neutral hydrogen \cite{pinetti2019} and the cosmic voids to trace the large scale structure of the Universe \cite{Arcari:2022zul}.

This paper aims at investigating two improved statistical techniques that could potentially boost the sensitivity of the cross-correlation technique on the dark matter signal. The first method, inspired by the results of Refs. \citep{Alonso_Wiener_GW,Urban_Alonso_Camera_Wiener}, is based on the Wiener filter formalism \citep{Wiener_1949,Rybicky_Press_1992}, and relies on the identification of an optimal weighting scheme for the redshift distribution of observed galaxies to be cross-correlated with the UGRB. We derive the weighting scheme and identify the optimal strategy, by showing that it can indeed increase the signal-to-noise ratio of the cross-correlation between gamma-rays and galaxy catalogs for the determination of the presence of a dark matter signal in the UGRB. The analysis is focussed on a low-redshift spectroscopic catalog modeled on the 2MRS galaxy survey \citep{Huchra_2MRS_2012}. The reason for the adoption of the 2MRS catalog is based on several reasons. First of all, 2MRS is spectroscopic, which makes it especially suitable for the filtering technique that we will be presenting, based on the adoption of a redshift-dependent weighting scheme. Secondly, 2MRS contains galaxies at low redshift, which makes it valuable for its high overlap with the gamma-ray signal originating from annihilating DM, which is more concentrated at low redshift, while the emission from unresolved astrophysical sources, which represent the background for the DM signal, peaks at higher redshift \cite{camera2013novel,camera2015tomographic,fornengo2014}. Finally, 2MASS and 2MRS have been employed in several studies on cross-correlation techniques \cite{xia2011,Ando_2MRS_2014,xia2015tomography,regis2015particle,cuoco2015dark,cuoco2017tomographic,paopiamsap2024constraints,Pinetti:2025qpi}, and this allows us a direct comparison of our proposed techniques with the existing literature.
The second method relies on taking advantage of a full multi-tracer information. In this case, the statistical information brought by the cross-correlation between gamma-rays and galaxies is combined  with the additional information encoded in the auto-correlation signals of the galaxies and of the auto-correlation signal of the gamma-rays field. We will show that the inclusion of the full multi-tracer information can improve the sensitivity to dark matter, even though only for the lower end of the WIMP mass range.

The paper is structured as follows: the main formalism for cross-correlations is introduced in Sec. \ref{sec:formalism_power_spectrum}. Sec. \ref{sec:Wiener_filter} discusses the Wiener filter and the case for optimal weighting. Sec. \ref{sec:analysis} discusses the analysis methodology, introducing also the multi-tracer technique. Sec. \ref{sec:results} presents our results and finally Sec. \ref{sec:conclusions} draws our conclusions. Further details on the formalism are collected in the Appendixes. All our calculations are embedded within a $\Lambda$CDM model using the parameters of the Planck 2018 release \citep{Planck_2018}.

\section{Cross-correlation angular power spectra}
\label{sec:formalism_power_spectrum}

In this Section we introduce the basic elements that enter the definition of the auto- and cross-correlation angular power spectra (APS) and of their (co-)variances. The main observable under scrutiny in this paper is the cross-correlation APS between a cosmological background radiation field emitted by dark matter structures in the Universe (specifically, the unresolved gamma-ray background -- UGRB) and a tracer of the dark matter distribution in the Universe (specifically, galaxies as identified in galaxy catalogs). The aim of the paper is to develop a filtering technique based on optimal weighting, as an attempt in improving the extraction of a dark matter signal. The filter will be applied to the galaxy redshift distribution (a quantity that is observationally available) and will be modeled in terms of the expected redshift distribution of the gamma-ray emission, 
that can be theoretically modeled but that is not available observationally, since only the gamma-ray flux integrated over distance is available to a telescope. The scope of the analysis is to understand whether and under which circumstances a filter can be successfully applied to improve the signal-to-noise ratio of the cross-correlation observable and to forecast the bounds on the particle physics properties of the dark matter particle (namely, its mass and annihilation cross section) that can be obtained.  Since the full expressions for the cross-correlations APS contain a large number of ingredients, for the sake of readability we introduce here only the essential elements useful to understand the logic of the proposed methodology, and refer the reader to Appendices \ref{app:dark_matter}, \ref{app:astro_sources}, \ref{app:galaxies} and \ref{app:3D_PS} for all the details.

Given two observables $A$ and $B$, their auto- or cross-correlation angular power spectra measures the correlation of their harmonic-expansion coefficients. For a statistically homogeneous field on the sphere, the spectrum depends only on the angular multipole $\ell$, corresponding to the angular scale separating any two points on the sky, i.e.:

\begin{equation}
    \langle A_{\ell m}\,B_{\ell'm'}\rangle=\delta^{\rm K}_{\ell\ell'}\,\delta^{\rm K}_{mm'}\,C^{AB}_\ell\;,
\end{equation}
with angle brackets denoting ensemble average and \(\delta^{\rm K}\) being the Kr\"onecker-delta symbol.
In the so-called Limber approximation \citep{limber1953,limber1992,limber1998} 
valid for $\ell\gg1$, the APS reads:

\begin{equation}
 C_{\ell}^{AB} = \int \frac{\de\chi}{\chi^2} \, W_A(\chi)\,W_B(\chi)\,P_{AB}\left(k=\frac{\ell+1/2}{\chi}, \chi\right),
\label{eq:APS}
\end{equation}
where $k$ denotes the wavenumber of a Fourier space decomposition, $\chi$ is the radial comoving distance, related to redshift $z$ in a flat cosmology through the relation $d\chi = c\,dz/H(z)$ with $H(z)$ being the expansion rate of the Universe.

In Eq. (\ref{eq:APS}) $P_{AB}(k,z)$ denote the 3D cross-correlation power spectra of the fluctuations of the $A,B$ fields, which, under the assumption of Gaussian fluctuations, are defined as:

\begin{equation}
\langle \tilde f_A({\bm k},z) \tilde f_B({\bm k},z) \rangle = (2\pi)^2 \delta_D^3({\bm k} - {\bm k}') P_{AB}(k,z)
\label{eq:3DPS}
\end{equation}
where $f_A = (g_A -\langle g \rangle) /\langle g \rangle$ are the fluctuations of the field $g_A$ around its mean, normalized to the mean. $P_{AB}(k,z)$ quantifies the co-variance of the two fluctuation-fields $f_{A,B}$ at spatial scale $k^{-1}$ and redshift $z$. Since our analysis requires modeling of medium and small-scale clustering of both dark matter halos and astrophysical sources, linear perturbation theory is insufficient to properly determine the clustering: we therefore resort to the halo model formalism \citep{cooray2002}, \citep{Asgari_halo_model_2023}. In this framework, the power spectrum $P_{AB}(k,z)$ can be separated into two terms, usually referred to as the 1-halo and the 2-halo term, i.e.\ :
\begin{equation}
P_{AB}(k) = P_{AB}^{\rm 1h} (k) + P_{AB}^{\rm 2h}(k).
\label{eq:pk}
\end{equation}
The 1-halo term refers to the case where the contribution to the correlation between the two fields $A,B$ comes from the same halo, while the 2-halo term describes contributions from two different halos. On large scales, the 2-halo term becomes proportional to the linear power spectrum, properly weighted by bias factors that relate the fields $A,B$ to the underlying dark matter field. On small scales, typically the 1-halo term provides the dominant contribution. Explicit expressions and dependencies on the physical modeling are given in Appendices \ref{app:dark_matter}, \ref{app:astro_sources}, \ref{app:galaxies} and \ref{app:3D_PS}. In all of the above, when $A=B$ we are dealing with the auto-correlation of a single field.

In Eq. (\ref{eq:APS}), $W_{A,B}(\chi)$ are the so-called window-functions and represent a kernel that expresses the redshift dependence of the intensity field of the observable $A$ or $B$. In our case, the observables are the galaxy number counts and the gamma-ray emission field, which is in turn composed by a contribution from dark matter annihilation in cosmic structures and in the emission from astrophysical sources located in the same structures. The latter represents an irreducible background for the dark matter signal. For the gamma-ray emissions (which in our case are represented by the DM emission and  by the inevitable emission from unresolved astrophysical sources) the window functions are normalized to the average intensity of the corresponding flux:
\begin{equation}
\langle I_A \rangle = \int d \chi\  W_A(\chi).
\label{eq:av_intensity}
\end{equation}
For annihilating dark matter the expression is:
\begin{equation}
W_{\rm DM}(E, \chi) = \frac{dW_{\rm DM}}{dEd\chi}= \frac{1}{4 \pi} \, \frac{\langle \sigma v \rangle}{2} \Delta^2(z) \, \left (\frac{\Omega_{\rm DM} \,\rho_{\rm c} }{m_{\chi}} \right)^2 (1+z)^3 \frac{d N}{d E}[(1+z)E] {\rm e}^{-\tau \left(E,  z \right)}, \label{eqn:Wann}
\end{equation}
where $\langle \sigma v \rangle$ is the annihilation rate of a pair of dark matter particles, $m_\chi$ the dark matter particle mass, $E$ the photon energy,
$\Omega_{\rm DM}$ the dark matter density parameter, $\rho_c$ the critical density of the Universe at present time, $\Delta(z)^2$ the clumping factor, $dN/dE$ the energy spectrum of photons produced in an annihilation event in the rest-frame of the process, and $\tau(E,z)$ the optical depth for gamma-ray absorption along the line of sight, caused by pair production on the extra-galactic background light emitted by galaxies in the ultraviolet, optical, and infrared bands, for which we adopt the results from Ref. \cite{Franceschini_2008}. Full details of the modeling are reported in Appendix \ref{app:dark_matter}. The dark matter particle mass $m_\chi$ and the annihilation rate $\langle \sigma v \rangle$ are the particle physics parameters that modulate the size of the signal, while the energy spectra determine its energy behavior and in turn depend on the type of annihilation products available to the dark matter particle. We will provide forecasts on the ability of the ``improved'' cross-correlation technique to set bounds on $\langle \sigma v \rangle$ as a function of $m_\chi$, for a dark matter particle annihilating into a pair of $b\bar b$ quarks , which is often the dominant annihilation channel for dark matter particles composed by weakly interacting massive particles (WIMP). The energy spectra at the source are modeled as in Ref. \citep{Arina_Cosmix_2023}.

\begin{table*}[t!]
\centering
\begin{tabular}{|ccccccc|}
\hline Bin & \begin{tabular}{c}
$E_{\min}$ \\
$(\mathrm{GeV})$
\end{tabular} & \begin{tabular}{c}
$E_{\max}$ \\
$(\mathrm{GeV})$
\end{tabular} & \begin{tabular}{c}
$\nn^\gamma$ \\
$\left(\mathrm{cm}^{-4} \mathrm{~s}^{-2} \mathrm{sr}^{-1}\right)$
\end{tabular} & $f^\gamma_{\text {sky }}$ & \begin{tabular}{c}
$\sigma_{0}^{\text {Fermi }}$ \\
$(\mathrm{deg})$
\end{tabular} & \begin{tabular}{c}
$E_{\mathrm{b}}$ \\
$(\mathrm{GeV})$
\end{tabular} \\
\hline 1 & 0.5 & 1.0 & $1.056 \times 10^{-17}$ & 0.134 & 0.87 & 0.71 \\
2 & 1.0 & 1.7 & $3.548 \times 10^{-18}$ & 0.184 & 0.50 & 1.30 \\
3 & 1.7 & 2.8 & $1.375 \times 10^{-18}$ & 0.398 & 0.33 & 2.18 \\
4 & 2.8 & 4.8 & $8.324 \times 10^{-19}$ & 0.482 & 0.22 & 3.67 \\
5 & 4.8 & 8.3 & $3.904 \times 10^{-19}$ & 0.549 & 0.15 & 6.31 \\
6 & 8.3 & 14.5 & $1.768 \times 10^{-19}$ & 0.574 & 0.11 & 11.0 \\
7 & 14.5 & 22.9 & $6.899 \times 10^{-20}$ & 0.574 & 0.09 & 18.2 \\
8 & 22.9 & 39.8 & $3.895 \times 10^{-20}$ & 0.574 & 0.07 & 30.2 \\
9 & 39.8 & 69.2 & $1.576 \times 10^{-20}$ & 0.574 & 0.07 & 52.5 \\
10 & 69.2 & 120.2 & $6.205 \times 10^{-21}$ & 0.574 & 0.06 & 91.2 \\
11 & 120.2 & 331.1 & $3.287 \times 10^{-21}$ & 0.597 & 0.06 & 199.5 \\
12 & 331.1 & 1000 & $5.094 \times 10^{-22}$ & 0.597 & 0.06 & 575.4 \\
\hline
\end{tabular}    
\caption{Gamma-ray energy bins used in our analysis, adherent with 8 years of data taking from \Fermi\ Pass 8 \cite{ackermann2018}. $N^\gamma$ is the measured auto-correlation noise (which, in our analysis, we rescale down by a factor of 2.5 in order to account for a detector exposure of approximately 20 years), $f^\gamma_{\text {sky }}$ the observed fraction of the sky outside the combined Galactic and point-source masks and $\sigma_0^{\text {Fermi }}$ the $68 \%$ containment angle of the PSF and $E_b=\sqrt{E_{\min } E_{\max }}$ the geometric center of each energy bin.}
\label{tab:Fermi_energy_bins}
\end{table*}

\begin{figure}[t!]
    \centering
    \includegraphics[width=1\linewidth]{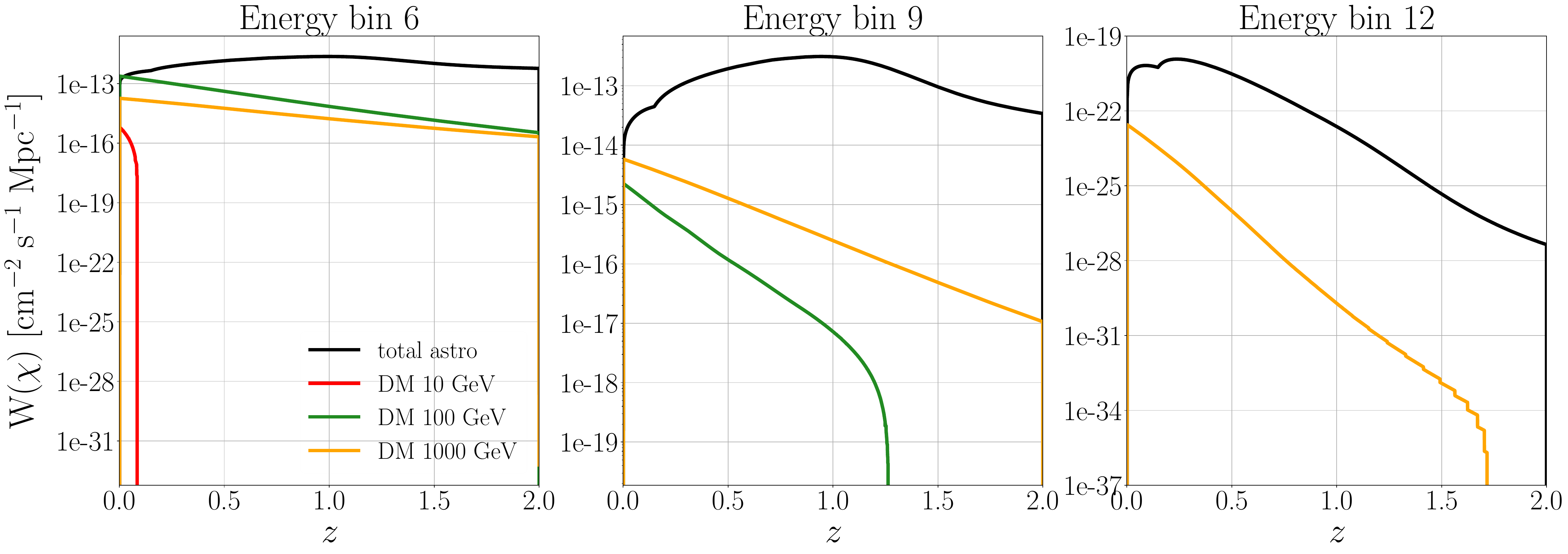}
    \caption{Gamma-ray window functions $W_\gamma(z)$ as a function of redshift for the total emission (red lines) from unresolved astrophysical sources (blazars, misaligned active galactic nuclei, flat spectrum radio quasars and star forming galaxies), together with a few representative examples from dark matter emission in the $b\bar b$ channel, namely: $\mdm = 10$ GeV (red), $\mdm = 100$ GeV (green) and $\mdm = 1000$ GeV (blue). In all cases the cross section is set at its ``thermal'' value $\svth = 3\times 10^{-26}$ cm$^{-3}$ s$^{-1}$. The three panels refer to the gamma-ray emission in three different energy bins; from left to right: number 6, 9 and 12 of Table \ref{tab:Fermi_energy_bins}. }
    \label{fig:UGRB_windows}
\end{figure}

\begin{figure}[t!]
    \centering
    \includegraphics[width=1\linewidth]{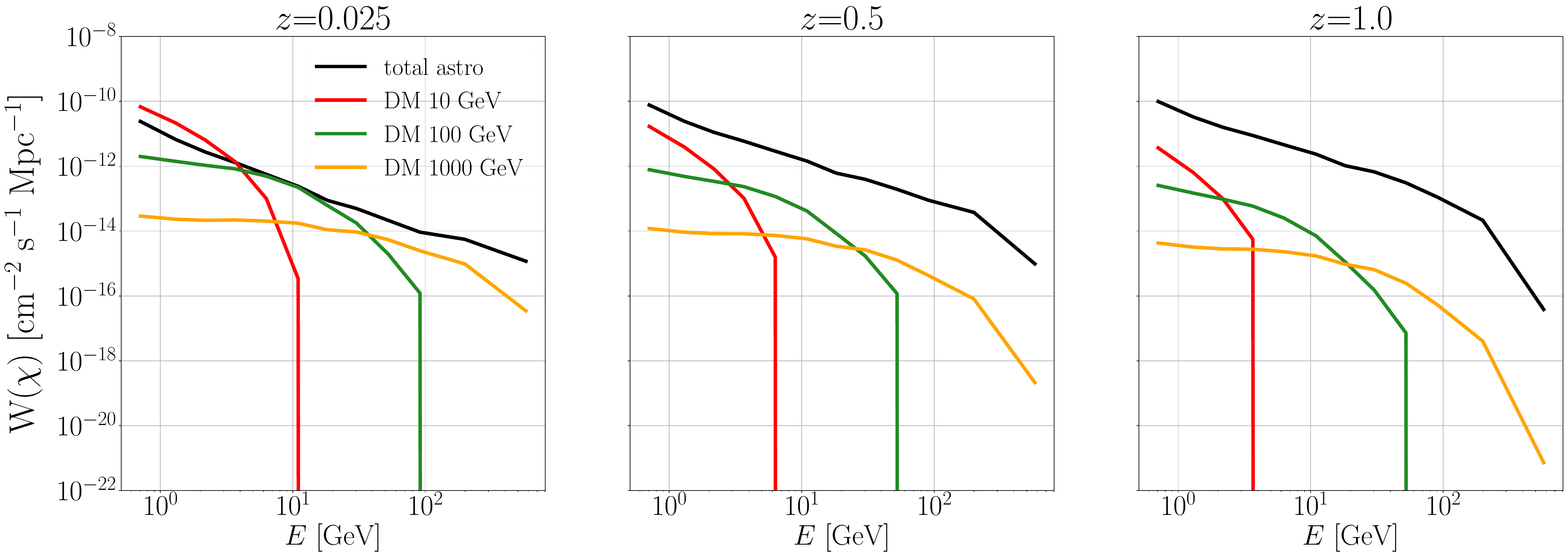}
    \caption{Gamma-ray window functions $W_\gamma(E)$ as a function of energy for the total emission (red lines) from unresolved astrophysical sources (blazars, misaligned active galactic nuclei, flat spectrum radio quasars and star forming galaxies), together with a few representative examples from dark matter emission in the $b\bar b$ channel, namely: $\mdm = 10$ GeV (red), $\mdm = 100$ GeV (green) and $\mdm = 1000$ GeV (blue). In all cases the cross section is set at its ``thermal'' value $\svth = 3\times 10^{-26}$ cm$^{-3}$ s$^{-1}$. The three panels refer to the gamma-ray emission in three different redshift bins; from left to right: $z = 0.025$, $z = 0.5$, $z = 1$. }
    \label{fig:UGRB_windowsE}
\end{figure}

As astrophysical gamma-ray sources, we consider four classes: blazars (BLA), misaligned active galactic nuclei (mAGN), flat spectrum radio quasars (FSRQ) and star forming galaxies (SFG). The window function for each source class has the form:
\begin{equation}
     W_{\rm astro}(E, z) =  \frac{dW_{\rm astro}}{dEd\chi}= \left(\frac{d_L(z)}{1+z} \right)^2   \int_{\mathcal{L}_{\rm min}}^{\mathcal{L}_{\rm max}(z)} d \mathcal{L} \, \frac{d F}{d E}(E,\mathcal{L},z) \, \phi(\mathcal{L},z) e^{-\tau(E,z)}, \label{eq:wastro}
\end{equation}
where $\mathcal{L}$ is the source luminosity, $d_L(z) = (1+z)\chi(z)$ the luminosity distance, $\phi(\mathcal{L},z)$ the gamma-ray luminosity function and $dF/dE$ the spectral energy distribution (SED) of the sources. The window function collects all sources with luminosity comprised between a minimal luminosity $\mathcal{L}_{\rm min}$, which depends on the intrinsic properties of the source class, and a maximal luminosity $\mathcal{L}_{\rm max}(z)=\min[\mathcal{L}_{\rm max}^{\rm intr}, \mathcal{L}_{\rm sens}(z)]$ where $\mathcal{L}_{\rm max}^{\rm intr}$ again is an intrinsic property of the source class and $\mathcal{L}_{\rm sens}(z)$ is related to the sensitivity of the detector: since we are considering {\it unresolved} sources, we need to account only for those sources, located at redshift $z$, that contribute a photon flux at the detector below the detector sensitivity $F_{\rm sens}$. The full details of the modeling are reported in Appendix \ref{app:astro_sources}, for all the four source categories considered in our analysis. The astrophysical sources represent the irreducible background for the dark matter gamma-ray signal, and the parameters entering their modeling will be fixed as reported in Appendix \ref{app:astro_sources}. A few examples of gamma-ray window functions as a function of redshift and for some representative energy bins are shown in Fig. \ref{fig:UGRB_windows}
and as a function of energy for a few representative redshift values in Fig. \ref{fig:UGRB_windowsE}.

\begin{figure}[t!]
    \centering
    \includegraphics[width=1\linewidth]{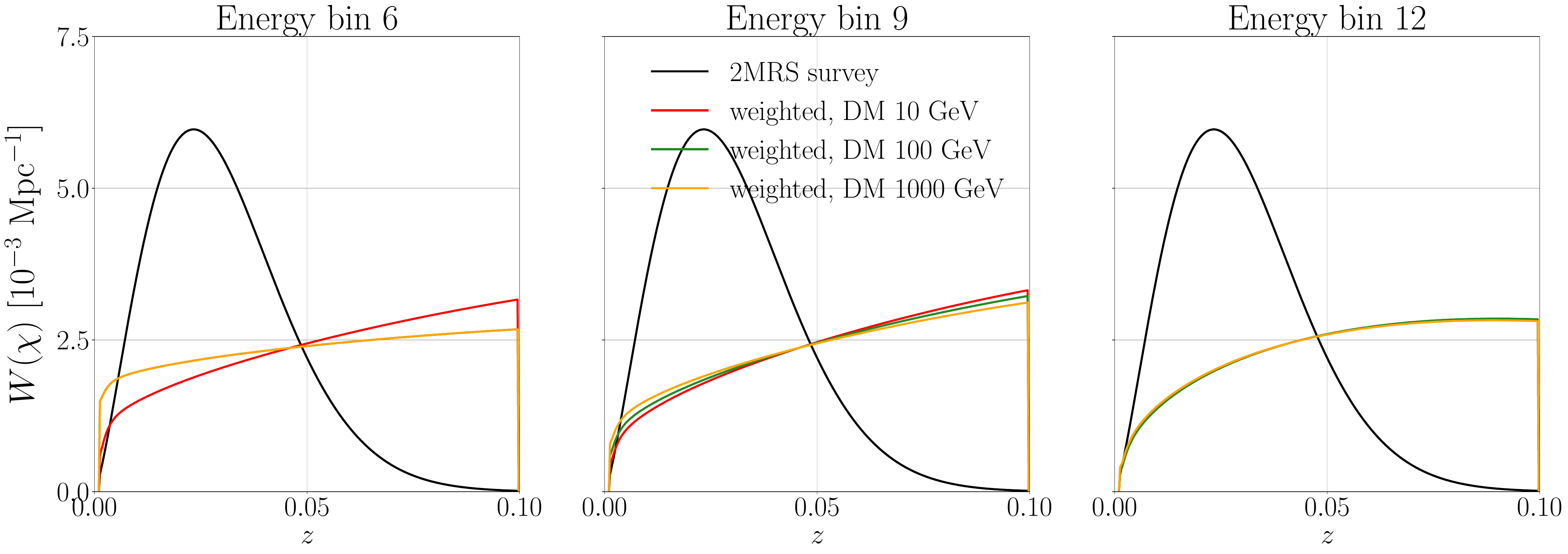}
    \caption{Galaxy survey window functions vs redshift. The black lines shows the 2MRS window function $W_g(\chi)$ without applying weights (standard case). The remaining lines
    show the 2MRS window functions $W_{wg}(\chi)$ with the weights applied. The colors differentiate the dark matter mass (red is for $m_{\rm DM} = 10$ GeV, green for $m_{\rm DM} = 100$ GeV and yellow for $m_{\rm DM} = 1000$ GeV -- in all cases the cross section is set at its ``thermal'' value $\sv = 3\times 10^{-26}$ cm$^{-3}$ s$^{-1}$) while the three panels, from left to right, refer to the \Fermi\ energy bins number 6, 9 and 12. The unweighted window functions are clearly the same in each panel. }
    \label{fig:galaxy_windows}
\end{figure}

Finally, the window function of a galaxy catalog depends on the comoving number density of galaxies $\bar{n}_{g}(\chi)$ as:
\begin{equation}
    W_{g}( \chi ) = \frac{\chi^2 \, \bar{n}_{g}(\chi)}{\left[\int d\chi \, \chi^2  \, \bar{n}_{g}(\chi) \right]}. \label{eq:gal_window_function0}
\end{equation}
In this way, the galaxy window function is normalized:
\begin{equation}
\int d\chi\,  W_g(\chi) = 1, 
\end{equation}
with 
\begin{equation}
\int d\chi \, \chi^2  \, \bar{n}_{g}(\chi) = {\bar N}_{\Omega, g}, 
\label{eq:number_density}
\end{equation}
${\bar N}_{\Omega, g}$ being the number density of galaxies per steradian. We will adopt a window function modeled on the 2MRS galaxy catalog \citep{Huchra_2MRS_2012}. Details are given in Appendix \ref{app:galaxies}: here we just recall that the 2MRS catalog contains 43\,812 galaxies in the redshift range $(0.0012, 0.1)$. The average angular number density of the survey is 3920 galaxies/sr$^{-2}$. The redshift behaviour of the 2MRS window function is shown in Fig. \ref{fig:galaxy_windows} as a black solid curve.

The filtering technique discussed in the next Section attempts at deriving optimal weights $w(\chi)$ to be applied to the galaxy kernel, in order to maximize the signal-to-noise ratio for the cross-correlation observable $C^{\gamma g}_\ell$. In this situation, the (normalized) weighted galaxy window function becomes:
\begin{equation}
    W_{wg}( \chi ) = \frac{\chi^2 \ w(\chi) \ \bar{n}_{g}(\chi)}{\left[\int d \chi  \ \chi^2 \ w(\chi) \ \bar{n}_{g}(\chi) \right]}. 
    \label{eq:gal_window_function}
\end{equation}
We anticipate that the optimal weight $w(\chi)$ will be provided by the gamma-ray window function $W_\gamma(z)$, thus ``aligning'' the galaxy distribution used to determine the cross-correlation APS to a theoretically-expected redshift distribution of the gamma-ray emission. This procedure can be operationally performed on data: the observationally available redshift distribution of galaxies can be modified through the adoption of a filter function $w(\chi)$ which traces a theoretical model for $W_\gamma(\chi)$ and cross-correlated with the observed angular gamma-ray map. This procedure can be repeated for different models of gamma-ray emission from dark matter (and astrophysical sources, as well) and a figure of merit like the signal-to-noise ratio can then be used to determine the statistical significance of the agreement, thus inferring information of the underlying theoretical models. We will investigate and quantify whether the filtering technique has potentiality to improve the bounds on the dark matter annihilation rate as a function of the dark matter mass, and identify the best strategy of filtering. The derivation of the optimal filter is outlined in Section \ref{sec:Wiener_filter}. 

The covariance matrix of the harmonic-space power spectrum $C_\ell^{AB}$ reads 
\begin{equation}
\Gamma^{AB}_{ar\ell,bs\ell'} = \frac{ \delta^{\rm K}_{\ell\ell'}}{(2\,\ell+1)\,f_{\rm sky}\,\Delta \ell}\,\left[\left(C_\ell^{ar} + \nn^{ar}\right)\,\left(C_\ell^{bs} + \nn^{bs}\right)+ \left(C_\ell^{ab} + \nn^{ab}\right)\,\left(C_\ell^{rs}+ \nn^{rs}\right) \right]\;,\label{eq:covCl}
\end{equation}
where we have used the Gaussian approximation, which makes the covariance diagonal in harmonic space ($\delta^{\rm K}$ is the Kronecker delta symbol), the $C_\ell^{AB}$'s denote auto- or cross-spectra and $\cal N$ the noise terms, which are independent of $\ell$. $\Delta \ell$ is the width of the $\ell$-bin and $f_{\rm sky}$ the sky-coverage fraction. Indices $a,b$ refer to features of field $A$, while $r,s$ to features of field $B$. By features we mean the energy-bin for the gamma-ray field and the redshift-bin for the galaxy catalog. Note that in the absence of redshift binning, as in the present analysis, we simply have \(r,s=1\).


The sky-coverage fraction for the 2MRS galaxy catalog is taken as $f^g_{\rm sky} = 0.877$ \citep{Ando_2MRS_2018}. For \Fermi, the sky-coverage typically depends on the energy bin, since it is related to the portion of the sky which is masked either because dominated by the Galactic foreground emission or by the presence of resolved astrophysical sources. To be definite, we have based our analysis on the \Fermi\ specifications of Ref. \cite{ackermann2018}, reported in Table \ref{tab:Fermi_energy_bins}. We have therefore adopted the same energy bins of Ref. \cite{ackermann2018} and the same energy-dependent $f^{\gamma}_{{\rm sky}, i}$ ($i$ labels the energy bin). For each entry in the covariance matrix $\Gamma^{AB}_{ar\ell,bs\ell'}$, we have then adopted the criteria: $f_{\rm sky} = \min[f^g_{\rm sky},f^{\gamma}_{{\rm sky},i}]$ for cross-correlations ($A = {\rm galaxies}$, $B = {\rm gamma\; rays}$) and $f_{\rm sky} = \min[f^{\gamma}_{{\rm sky},i}, f^{\gamma}_{{\rm sky},j}]$ ($A,B = {\rm gamma\; rays}$ in energy bins $i$ and $j$) for gamma-ray auto-correlations.

The noise terms for gamma rays depend on energy but can be considered as diagonal in this variable, since the number counts of photons in different energy bins are independent:
\begin{equation}
\nn^{\gamma\gamma}_{ij} = \delta^K_{ij} \nn^\gamma \label{eq:noise_auto_gamma}
\end{equation}
The measured values for $\nn^\gamma$ for 8 years of data taking from \Fermi\ Pass 8 obtained in Ref. \cite{ackermann2018} are reported in Table \ref{tab:Fermi_energy_bins}. In order to adapt our forecasts to 20 years of operation, since the noise scales approximately with the inverse of the exposure \cite{pinetti2019,Arcari:2022zul} we have rescaled the noise values of Table \ref{tab:Fermi_energy_bins} by a factor of 2.5.

The noise for galaxies is obtained as:
\begin{equation}
\nn^{gg} = \frac{4\pi f^g_{\rm sky}}{N_{\rm survey}}
\label{eq:noise_galaxies}
\end{equation}
where  $N_{\rm{survey}}$ is the number of galaxies observed in the given survey and:
\begin{equation}
    \nn^{g\gamma}_i = 0 \label{eq:noise_cross_gal_gamma}
\end{equation}

When the filter $w(\chi)$ is applied, the situation becomes more complex. We have already anticipated that the filter is applied to the galaxy distribution and it depends on the gamma-ray window functions, which are energy dependent. In this case, we have different galaxy noise terms for each \Fermi\ energy bin. In this case the galaxy noise has to be accordingly modified as:
\begin{equation}
  \mathcal{N}^{g g}_{ij} = \frac{\int \de\chi\;\chi^2\,w_i(\chi)w_j(\chi)\,\bar{n}_\gal(\chi)}{\left[\int \de\chi\; \chi^2\,w_i(\chi)\,\bar{n}_\gal(\chi)\right] \left[\int \de\chi\; \chi^2\,w_j(\chi)\,\bar{n}_\gal(\chi)\right]} \,. \label{eq:noise_gopt}
\end{equation}
which, taking into account Eq.(\ref{eq:number_density}), reproduces the standard case of Eq. (\ref{eq:noise_galaxies}) for $w(\chi)= 1$, i.e. when no filter is applied.

All theoretical terms involving \g-rays need corrections due to the energy-dependent \Fermi\ point spread function (PSF), whose expression is provided in Appendix \ref{app:Fermi_PSF}. This means that, in angular space, correlation functions of the type $C_\ell^{g\gamma_i}$ get multiplied by the corresponding \Fermi\ PSF in angular space $B_{\ell, i}$ for that specific energy bin (i.e. $C_{\ell, i}^{g\gamma} \rightarrow C_{\ell, i}^{g\gamma} B_{\ell, i}$) and correlation functions of the type $C_{\ell,ij}^{\gamma\gamma}$ receive twice the beam correction (i.e. $C_{\ell,ij}^{\gamma\gamma} \rightarrow C_{\ell,ij}^{\gamma\gamma} B_{\ell,i} B_{\ell,j}$). Noise terms instead, being related to Poisson fluctuations in photon number counts, are not affected by the angular resolution and therefore they are not corrected for the beam. For terms involving galaxies instead, we have not applied a beam function, since the angular resolution of galaxy catalogs is much better than the one for gamma-rays and therefore it starts to be relevant for multipoles which are already significantly affected by the \Fermi\ PSF. In fact, it is the \Fermi\ PSF that will determine that our analysis will be meaningful at most up to multipoles $\ell \sim 1000$, corresponding to angular scales of the order of $0.1^\circ-0.2^\circ$, much larger than the typical angular resolution of galaxy catalogs of the order of the arcsec.

\begin{figure}[t!]
    \centering
    \includegraphics[width=1\linewidth]{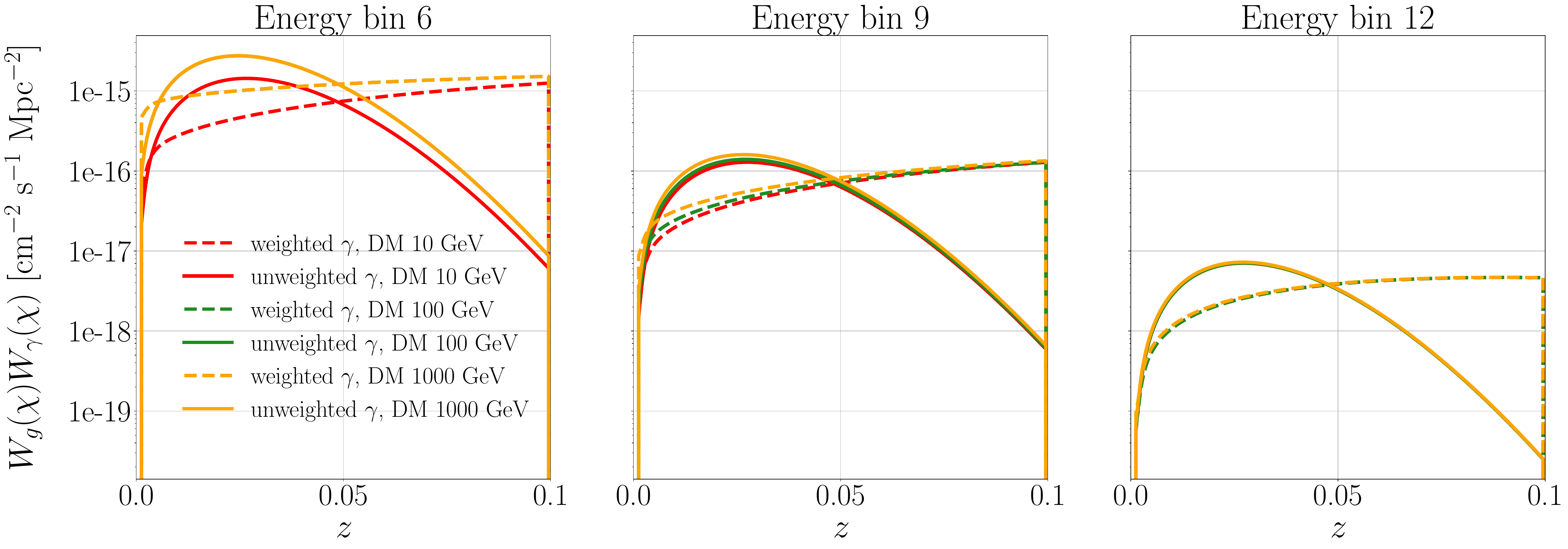}
    \caption{Some representative cases of the product between the galaxy window function $W_g(\chi)$ and the gamma-rays window function $W_\gamma(\chi)$, in the unweighted (standard) case (solid lines) and when weights are applied (dashed lines). In all cases, the gamma-ray window function adds the astrophysical sources contribution to a DM signal, which refers to: $m_{\rm DM} = 10$ GeV (red lines), $m_{\rm DM} = 100$ GeV (green) and $m_{\rm DM} = 1000$ GeV (yellow). In all cases the cross section is set at its ``thermal'' value $\sv = 3\times 10^{-26}$ cm$^{-3}$ s$^{-1}$. From left to right, the panels refer to the \Fermi\ energy bins number 6, 9 and 12. The quantity $W_g(\chi)$ $W_\gamma(\chi)$ determines the cross-correlation signal, as in Eq. (\ref{eq:APS}). }
    \label{fig:kernel_products}
\end{figure}

\section{Wiener filter: the case for optimal weighting}
\label{sec:Wiener_filter}

We are now ready to introduce the weighting technique we will employ in our work, following the discussions presented in 
\citep{Alonso_Wiener_GW,Urban_Alonso_Camera_Wiener}. Let us consider the sky-maps of gamma-ray number counts and galaxies. Each pixel contains $N^p_\gamma$ photons and $N^p_g$ galaxies. Each galaxy has an associated redshift: therefore we can build a vector $\boldsymbol{N^p_g}$ containing the number of galaxies in each redshift bin. To lighten the notation, let us call $\boldsymbol{x} = (N^p_{g1}, \cdots, N^p_{gM})$ the vector containing the number of galaxies in the different redshift bins ($M$ is the number of those bins) and $y=N^p_\gamma$. Let us also suppress the energy dependence of the photon counts, since the same argument can be repeated for each energy bin. Notice that boldface symbols denote vector and matrices, while non-boldface symbols are scalars.

The number of photons can be related to the number of galaxies in the same pixel, by means of a functional form connecting the two observables, which in our case can be assumed to be linear:
\begin{equation}
    y = \sum_r w_r x_r + n = \boldsymbol{w}^{\sf T} \boldsymbol{x} + n
    \label{eq:lin_reg}  
\end{equation}
where $\boldsymbol{w}$ is a vector containing the coefficients (weights) of the linear regression model and with $n$ being the uncorrelated, Gaussian white noise corresponding to the residuals ($r$ labels redshift) \citep{Rybicky_Press_1992}. The linear relation assumption corresponds to the signal processing technique known as Wiener filter \citep{Wiener_1949}.

The optimal weights $\boldsymbol{w}$ are obtained by solving a minimization problem. Assuming Gaussian statistics, the probability for obtaining the photon counts $y$ given the galaxy counts vector $\boldsymbol{x}$ is \cite{Alonso_Wiener_GW}:
\begin{equation}
-2\log {\cal P}(y|\boldsymbol{x}) = \boldsymbol{z}^{\sf T} \;{\rm \bf Cov}^{-1}[\boldsymbol{z},\boldsymbol{z}] \;\boldsymbol{z} - \boldsymbol{x}^{\sf T}\; {\rm\bf Cov}^{-1}[\boldsymbol{x},\boldsymbol{x}] \;\boldsymbol{x}
\end{equation}
where $\boldsymbol{z} = (y, x_1, \cdots, x_M)$ and ${\rm \bf Cov}(\boldsymbol{a},\boldsymbol{b})$ is the covariance between two  observables $\boldsymbol{a}$ and $\boldsymbol{b}$ (here and elsewhere, $\log$ denotes the natural logarithm). Explicitly \cite{Alonso_Wiener_GW, Urban_Alonso_Camera_Wiener}:
\begin{equation}
{\rm \bf Cov}[\boldsymbol{z},\boldsymbol{z}] = 
\begin{pmatrix}
{\rm Cov}[y,y] & {\rm \bf Cov}^{T}[\boldsymbol{x},y]  \\
{\rm \bf Cov}[\boldsymbol{x},y] & {\rm \bf Cov}[\boldsymbol{x},\boldsymbol{x}]
\end{pmatrix}
\end{equation}
For the sake of clarity, let us point out that ${\rm \bf Cov}[\boldsymbol{z},\boldsymbol{z}]$ has dimension $(M+1) \times (M+1)$, with ${\rm Cov}[y,y]$ a scalar (dimension 1), ${\rm \bf Cov}[\boldsymbol{x},y]$ a vector with dimension $(M\times 1)$ and ${\rm \bf Cov}[\boldsymbol{x},\boldsymbol{x}]$ a matrix of dimension $M\times M$. 

The maximization of the likelihood with respect to the variable of interest $y$, obtained by imposing $\partial_y \log{\cal P}(y|\boldsymbol{x})=0$, gives the minimum variance estimator for $y$  \cite{Alonso_Wiener_GW, Urban_Alonso_Camera_Wiener}:
\begin{equation}
    y = {\rm\bf Cov}^{\sf T}[\boldsymbol{x},y] \; {\rm\bf Cov}[\boldsymbol{x},\boldsymbol{x}]\; \boldsymbol{x}
    \label{eq:lin_reg}  
\end{equation}
from which we can directly read the expression for the optimal filter \cite{Alonso_Wiener_GW, Urban_Alonso_Camera_Wiener}:
\begin{equation}
\boldsymbol{w} = {\rm\bf Cov}^{-1}[\boldsymbol{x},\boldsymbol{x}]\; {\rm\bf Cov}[\boldsymbol{x},y] 
\label{eq:Wiener}
\end{equation}
By expressing Eq. (\ref{eq:Wiener}) in components, and restoring the notation in terms of number of galaxies and number of photons:
\begin{equation}
w_r = \sum_s {\rm\bf Cov}^{-1}[N^p_{gr},N^p_{gs}] \; {\rm\bf Cov}[N^p_{gs},N^p_\gamma]
\end{equation}
Under the assumption of Poissonian statistics, it can be shown that the  covariance of two Poisson samples equals the number of events in the intersections of the two samples \cite{Alonso_Wiener_GW, Urban_Alonso_Camera_Wiener}. In this case:
\begin{eqnarray}
    &&{\rm\bf Cov}[N^p_{gr},N^p_{gs}] = \delta_{rs}\; N^p_{gr} \\
    &&{\rm\bf Cov}[N^p_{gr},N^p_\gamma] = N^p_{\gamma r} 
\end{eqnarray}
where in the last equation $N^p_{\gamma r}$ refers to the number of photons in the redshift bin $r$. From Eq. (\ref{eq:av_intensity}) and (\ref{eq:number_density}) we have:
\begin{eqnarray}
    &&{\rm\bf Cov}[N^p_{gr},N^p_{gs}] = \chi^2  \, \bar{n}_{g}(\chi)\; d\chi \\
    &&{\rm\bf Cov}[N^p_{gr},N^p_\gamma] = W_\gamma(\chi) \; d\chi 
\end{eqnarray}
where we have restored an explicit redshift dependence through the comoving distance $\chi$, and expressed as $d\chi$ the (small) comoving-distance bin which refers to the redshift bin selected by the $\delta_{rs}$.

The expression for the filter to be applied to the galaxy distribution is then:
\begin{equation}
    w(\chi) = \frac{W_\gamma(\chi)}{\chi^2\, \bar{n}_g(\chi)} \; [\Theta(\chi-\chi_{\rm min})-\Theta(\chi-\chi_{\rm max}) ]
    \label{eq:weight}
\end{equation}
where $\chi_{\rm min}$ and $\chi_{\rm max}$ refer to the boundaries of the redshift bins under consideration, when a tomographic approach is adopted. This, in turn, makes the galaxy weighted window function of Eq. (\ref{eq:gal_window_function}):
\begin{equation}
    W_{wg}(\chi) = \frac{W_\gamma(\chi)}{\int_{\chi_{\rm min}}^{\chi_{\rm max}} d\chi W_\gamma(\chi)} \; [\Theta(\chi-\chi_{\rm min})-\Theta(\chi-\chi_{\rm max}) ]
    \label{eq:Wgfilter}
\end{equation}

The filter aligns the galaxy kernel to a model of gamma-ray emission. We emphasize that the integral at denominator in Eq. (\ref{eq:Wgfilter}) is calculated on the support of the galaxy redshift bin under consideration. This modified kernel is then used in the measurement of the cross correlation with the observed gamma-ray map, from which a statistical figure-of-merit can be obtained (we will adopt a signal-to-noise ratio, as defined in the next Section). The distribution of the values of the figures-of-merit obtained by varying the underlying theoretical models can then be used to infer information on the theoretical inputs adopted in the filter, namely to test different ansatz on the sources (astrophysics and DM) composing the UGRB and their redshift distribution. Let us recall that, when the filter is applied, also the galaxy noises are modified, as specified in Eq. (\ref{eq:noise_gopt}).

Fig. \ref{fig:galaxy_windows} shows the window function $W_g(\chi)$ of the 2MRS galaxy catalog, both in its standard unweighted case, and for three representative cases $W_{wg}(\chi)$ where the filter refers to the DM emission only. Fig. \ref{fig:kernel_products} instead shows the product $W_g(\chi)W_\gamma(\chi)$ of the galaxies and gamma-ray kernels, which is the key quantity that determines the cross-correlation function $C_\ell^{\gamma g}$, as expressed in Eq. (\ref{eq:APS}), with a filter that refers to both DM and astrophysical emission.

\section{Wiener filters and multi-tracers for dark matter searches}

\label{sec:analysis}

In this Section, we test whether applying the filter can improve the ability of the cross-correlation technique to set bounds on the dark matter particle physical parameters, namely its mass $m_\chi$ and its self-annihilation cross section $\sv$. To this aim, we adopt, as a statistical figure-of-merit, a signal-to-noise ratio SNR defined as (notice that the definition below stands for the square of the SNR):
\begin{equation}
    \snr^2 = {\sum_{\ell \ell'}\, \sum_{E E'} \sum_{z,z'}\, {\bm{x}^{\bm{T}}_{\ell,E,z}}\, \bm{{\Gamma}}_{\ell\ell'EE' zz'}^{-1} \,\bm{x}_{\ell',E', z'}}  
\label{eq:SNR1}
\end{equation}
where $\ell,\ell'$ denote multipoles, $E,E'$ label the energy bins and $z, z'$ the redshift bins. In this expression, $\bm{x}_{\ell,E,z}$ represents the array containing the ``data'' vector at a given multipole $\ell$, energy bin $E$ and redshift bin $z$, while $\bm{\Gamma}_{\ell\ell'EE' zz'}$ is the covariance matrix. Considering the fact that the 2MRS catalog contains galaxies in a restricted redshift range (see Fig. \ref{fig:galaxy_windows} for the 2MRS window function), we will perform our analysis on a single redshift bin. In this case, the redshift label can be dropped. Moreover, the covariance matrix is taken as diagonal in multipole space (see Eq. (\ref{eq:covCl})). In this case, the signal-to-noise-expression simplifies to:
 \begin{equation}
    \snr^2 = {\sum_\ell {\sum_{E E'}\, {\bm{x}^{\sf T}_{\ell,E}} \,\bm{{\Gamma}}_{\ell EE'}^{-1}\, \bm{x}_{\ell,E'}}}  \; = \; \sum_{\ell = \ell_{\rm min}}^{\ell_{\rm max}} \snr^2_\ell
\label{eq:SNR2}
\end{equation}
where:
\begin{equation}
{\snr_\ell}^2 = {\sum_{E E'}\, {\bm{x}^{\sf T}_{\ell,E}} \,\bm{{\Gamma}}_{\ell EE'}^{-1}\, \bm{x}_{\ell,E'}}    
\label{eq:SNR_ell}
\end{equation}
is the SNR brought by each multipole, cumulated in energy. In Eq.~(\ref{eq:SNR2}), $\ell_{\rm min}$ will be set to 50, since when dealing with real data the foreground contamination due to the large-angle galactic emission somehow limits the possibility to access very low multipoles (even in the case of foreground subtractions), while 
$\ell_{\rm max}$ will be set to 1000, to conform with the angular size of the \Fermi\ PSF (see Appendix \ref{app:Fermi_PSF} and Table \ref{tab:Fermi_energy_bins}).

We perform three types of analysis: in the first case (case I), we use only the cross-correlation signal, i.e. $\bm{x}_\ell = \lbrace C_\ell^{\gamma g}\rbrace$; in the second case (case II), we add to the data vector also the autocorrelation of galaxies, i.e. $\bm{x}_\ell = \lbrace C_\ell^{\gamma g}, C_\ell^{g g}\rbrace$; in the third case (case III), we further add the gamma-ray autocorrelation $\bm{x}_\ell = \lbrace C_\ell^{\gamma g}, C_\ell^{g g}, C_\ell^{\gamma \gamma}\rbrace$. To be definite, we will call the last two cases ``multi-tracer'' analyses. It is a strategy proposed in the context of cosmological analyses of the large-scale structure 
\citep{2004MNRAS.347..645P,Multitracer_McDonald_Seljak_2009,Multitracer_Seljak_2009}, which exploits the fact that all cosmological observables are inherently correlated. In particular, when studying the clustering properties of biased tracers of the underlying matter density field, the multi-tracer technique allows us to mitigate the effect of cosmic variance \citep{Abramo_Leonard_2013_Multitracer,2014MNRAS.442.2511F,2015ApJ...812L..22F}. In the case of the present analysis, the multi-tracer effectively comes into play as the various contributions to the UGRB are related to different galaxy populations --- that is, different biased tracers of the large-scale structure. In our case, we add to the UGRB the galaxies of 2MRS as one additional tracer. 

The size of the data vector $\bm{x}_\ell = \lbrace C_\ell^{\gamma g}\rbrace$ is the number of the gamma-ray energy bins $N$ (12 in our analysis); the size of 
$\bm{x}_\ell = \lbrace C_\ell^{\gamma g}, C_\ell^{g g}\rbrace$ is $N + 1$, since we add the galaxy-galaxy autocorrelation observable for a single redshift bin;  finally, the size of $\bm{x}_\ell = \lbrace C_\ell^{\gamma g}, C_\ell^{g g}, C_\ell^{\gamma \gamma}\rbrace$ is $N + 1 + N(N+1)/2$, since we add the gamma-ray auto-correlation, including their ``internal'' cross-correlations among different energy bins. In the latter case of gamma-ray auto-correlation, the angular power spectrum $C_\ell^{\gamma \gamma}$ contains the correlations among all the gamma-ray sources under consideration: $C_\ell^{\gamma \gamma} = C_\ell^{DM \otimes DM} + \sum_i C_\ell^{DM \otimes {\rm astro}_i} + \sum_{i,j} C_\ell^{{\rm astro}_i \otimes {\rm astro}_j}$ with $i,j$ counting blazars, mAGN, FSRQ and SFG. 

A key ingredient in the determination of the SNR is the covariance matrix. The covariance matrix for the analysis of case I (cross-correlation $\gamma g$) can be schematically represented as:
\begin{equation}
{\cal A} =    
    \begin{pmatrix} 
   (\gamma_1 g , \gamma_1 g) & (\gamma_1 g ,\gamma_2 g) & \cdots & (\gamma_1 g, \gamma_{N} g)\\
    (\gamma_2 g ,\gamma_1 g) & (\gamma_2 g ,\gamma_2 g) & \cdots & (\gamma_2 g, \gamma_{N} g)\\
    \vdots & \vdots & \ddots & \vdots \\
    (\gamma_N g , \gamma_1 g) & (\gamma_N g ,\gamma_2 g) & \cdots & (\gamma_N g, \gamma_{N} g)
    \end{pmatrix}
    \label{cov:I}
\end{equation}
where each entry in the matrix refers to an element of the covariance matrix $\Gamma$ of Eq. (\ref{eq:covCl}).
In the above equation, the indices $i$ on $\gamma_i$ label the gamma-ray energy bins. Matrix $\cal A$ is symmetric and has dimension ${\rm dim} ({\cal A})= N \times N$.

{\renewcommand{\arraystretch}{1.2}
\begin{table*}[t!]
\centering
\begin{tabular}{|l|l|l|}
\hline 
\multicolumn{2}{|c|}{Strategy} & \multicolumn{1}{|c|}{Criterion}\\
\hline
No filter & {\it Unweighted} &  \\
\hline
With filter & {\it Astro+DM} &  $W_\gamma(\chi) = W^{\rm astro}_\gamma(\chi) + W^{\rm DM}_\gamma(\chi)$\\
~ & {\it DM-only} & $W_\gamma(\chi) = W^{\rm DM}_\gamma(\chi)$ $\oplus$ $E\leq m_\chi$ \\
\hline
\end{tabular}   
\caption{Strategies of analysis for the filtering technique, with the criterion adopted for each case. See full text for details. Each strategy is adopted for each combination of observables: gamma-rays $\otimes$ galaxies cross-correlation ($\gamma g$), multi-tracer that combines gamma-rays $\otimes$ galaxies cross-correlation with galaxies auto-correlation ($\gamma g$ + $gg$) and multi-tracer that combines gamma-rays $\otimes$ galaxies cross-correlation, galaxies auto-correlation and gamma rays auto-correlation ($\gamma g$ + $gg$ + $\gamma\gamma$).}
\label{tab:strategies}
\end{table*}
}

For the multi-tracer analysis of case II ($\gamma g + gg$) the covariance matrix contains ${\cal A}$ as a sub-matrix, and is enlarged by a column and a row to the following expression:
\begin{equation}
{\cal B}= 
\begin{pmatrix} 
    \begin{tabular}{cccc|c}
    \multicolumn{4}{c|}{${\cal A}$} & $(\gamma_1 g, g g)$ \\
    ~ & & & & $(\gamma_2 g, g g)$ \\
    ~ & & & & $\vdots$ \\
    ~ & & & & $(\gamma_N g, g g)$ \\
    \hline
    $(gg, \gamma_1 g)$ & $(gg, \gamma_2 g)$  & $\cdots$ & $(gg, \gamma_N g)$ & $(gg)$       
    \end{tabular}
\end{pmatrix}
\end{equation}
with a size of:
\begin{equation}
{\rm dim} ({\cal B})= 
\begin{pmatrix} 
    \begin{tabular}{c|c}
    $N \times N$ & $N \times 1$ \\
    \hline
    $1 \times N$ & $1$        
    \end{tabular}
\end{pmatrix}
\end{equation}

Finally, for the multi-tracer analysis of case III ($\gamma g + gg + \gamma\gamma$), the covariance matrix contains matrix $\cal B$ and is further enlarged to (we adopt here a more compact notation for the large additional matrix blocks; $i,a,b$ run from 1 to $N$):
\begin{equation}
{\cal C}= 
\begin{pmatrix} 
    \begin{tabular}{c|c|c}
     \multicolumn{2}{c|}{${\cal B}$}  & $(\gamma_i g, \gamma_a \gamma_b)_{[a<b]}$ \\
    \cline{3-3}
    \multicolumn{2}{c|}{~}  & $(g g, \gamma_a \gamma_b)_{[a<b]}$ \\
    \hline
     $(\gamma_a \gamma_b)_{[a<b]}, \gamma_i g)$ & $( \gamma_a \gamma_b)_{[a<b]} , g g)$ & $(\gamma_i \gamma_j, \gamma_a \gamma_b)_{[i<j , a<b]}$ \\    
    \end{tabular}
\end{pmatrix}
\label{eq:matrixC}
\end{equation}
where labels $a,b$ and $i,j$ refer to the energy bins. The additional blocks are matrices/vectors whose sizes are reported below:

\begin{equation}
{\rm dim} ({\cal C})= 
\begin{pmatrix} 
    \begin{tabular}{c|c|c}
    $N \times N$ & $N \times 1$ & $N \times N (N+1)/2$ \\
    \hline
    $1 \times N$ & $1$ & $1 \times N (N+1)/2$ \\
    \hline
    $N (N+1)/2 \times N$ & $N (N+1)/2 \times 1$ & $N (N+1)/2 \times N (N+1)/2$
    \end{tabular}
\end{pmatrix}
\label{eq:matrixC_dim}
\end{equation}

From the SNR, we determine bounds on the DM particle parameters (its mass $m_{\rm DM}$ and annihilation cross section $\sv$) by using a $\Delta \chi^2$ statistics:
\begin{equation}
    \Delta \chi^2(m_{\rm DM}, \sv)  = \snr^2_{\rm{DM + astro}}(m_{\rm DM}, \sv) - \snr^2_{\rm{ astro}}
    \label{eq:delta_chi_2}
\end{equation}
where $\snr_{\rm{ astro}}$ is the SNR calculated when the gamma-ray emission is due only to astrophysical sources (i.e. no dark matter emission is present), while $\snr_{\rm{DM + astro}}(m_\chi, \sv)$ is the SNR calculated when a contribution from dark matter is added to the gamma-ray UGRB. The forecasts on the bounds on $\sv$ than can be obtained with the different methods are calculated for each (fixed) $m_{\rm DM}$ by setting the criterion $\Delta \chi^2 = 4$.

\begin{figure}[t!]
    \centering
    \includegraphics[width=1\linewidth]{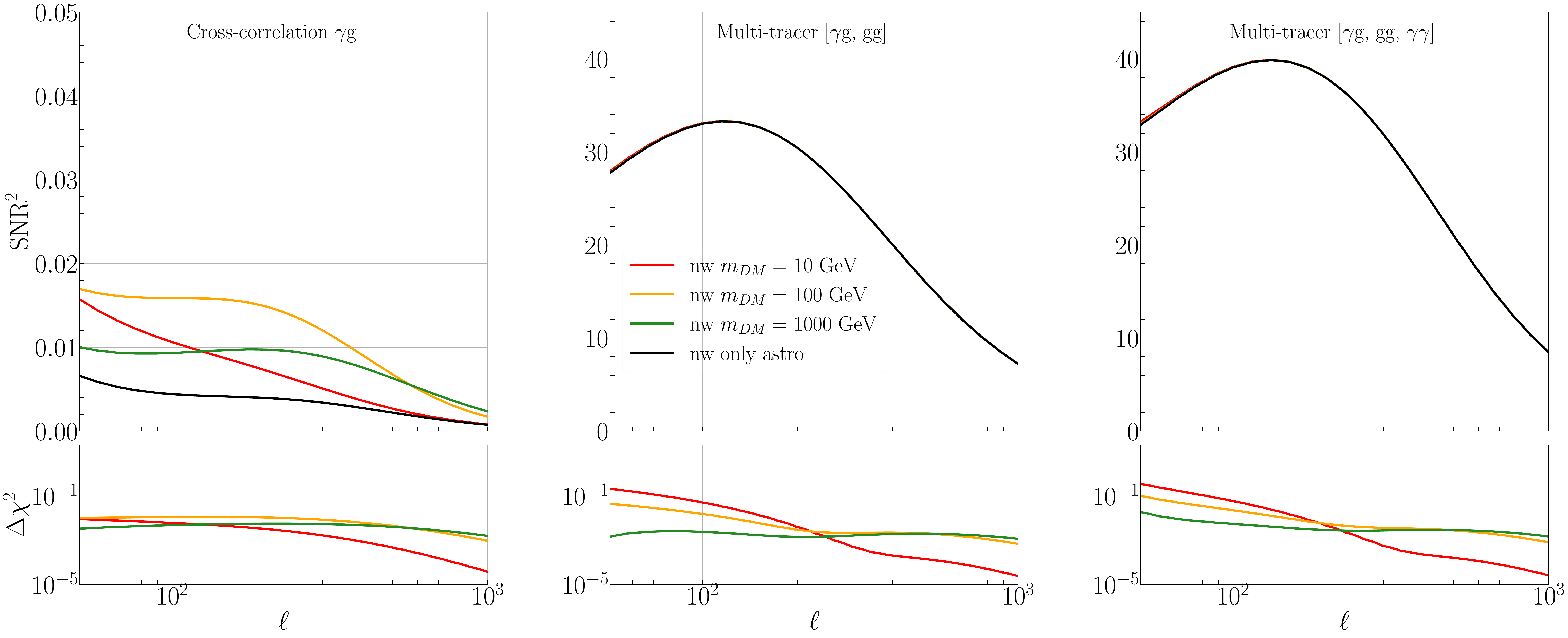}
    \caption{Contribution from each multipole to the SNR$^2$ (Eq. (\ref{eq:SNR_ell}), upper panels) and to the $\Delta\chi^2$ of Eq. (\ref{eq:delta_chi_2}) (lower panels) in the {\it unweighted} case. The left panels refer to the gamma-rays $\otimes$ galaxies cross-correlation $(\gamma g$) case, the central panels to the multi-tracer case that combines gamma-rays $\otimes$ galaxies cross-correlation with galaxies auto-correlation ($\gamma g$ + $gg$) and the right panels to the multi-tracer case that combines gamma-rays $\otimes$ galaxies cross-correlation, galaxies auto-correlation and gamma rays auto-correlation ($\gamma g$ + $gg$ + $\gamma\gamma$). In each upper panel, the black lines refer to gamma-ray emission from astrophysical sources only. The red, yellow and green lines stand for gamma-ray emission from astrophysical sources and dark matter, for three representative dark matter mass values: 10 GeV (red), 100 GeV (yellow) and 1000 GeV (green). The annihilation cross section $\sv$ for the three cases is 0.05, 0.5 and 5 times the ``thermal'' value $\svth = 3\times 10^{-26}$ cm$^{-3}$ s$^{-1}$, respectively.} 
    \label{fig:SNR_noweight}
\end{figure}

The SNR, and consequently the $\Delta \chi^2$, can be calculated with and without the filter applied. In this regard, we adopt two filtering (or weighting) strategies. In one case, called {\it astro+DM}, the gamma-ray window function adopted in Eq. (\ref{eq:weight}) contains the full gamma-ray emission, i.e. $W_\gamma(\chi) = W^{\rm astro}_\gamma(\chi) + W^{\rm DM}_\gamma(\chi)$, and the analysis is performed on the full energy range of \Fermi\ as outlined in Table \ref{tab:Fermi_energy_bins}. In the other case, called {\it DM-only}, we filter according only to the DM emission, i.e. $W_\gamma(\chi) = W^{\rm DM}_\gamma(\chi)$ and, for each DM model, we limit the gamma-ray energy range of the analysis to the interval where the non-relativistic DM annihilation actually produces gamma-ray, i.e. to energies such that $E \leq m_\chi$. To clarify: for the {\it DM-only} strategy, the astrophysical gamma-ray emission is present in the ``data'', but the filter of Eq. (\ref{eq:weight}) that is applied to obtain the galaxy weighted window function $W_{wg}(\chi)$ refers only to the dark matter emission, i.e. to $W^{\rm DM}_\gamma(\chi)$, which in turn is defined only for $E \leq m_\chi$: this implies that in this analysis we limit the energy range to conform to this limitation. In this case, for consistency and for not improperly penalizing the $\Delta\chi^2$, also the $\snr^2_{\rm{ astro}}$ is calculated for energies smaller than the DM mass (``capped'' $\snr^2_{\rm{ astro}}$). 

Table \ref{tab:strategies} summarizes the various types of strategies that we are investigating and whose results are presented and discussed in the next Section.

\section{Results and discussion}
\label{sec:results}

\begin{figure}[t!]
    \centering
    \includegraphics[width=1\linewidth]{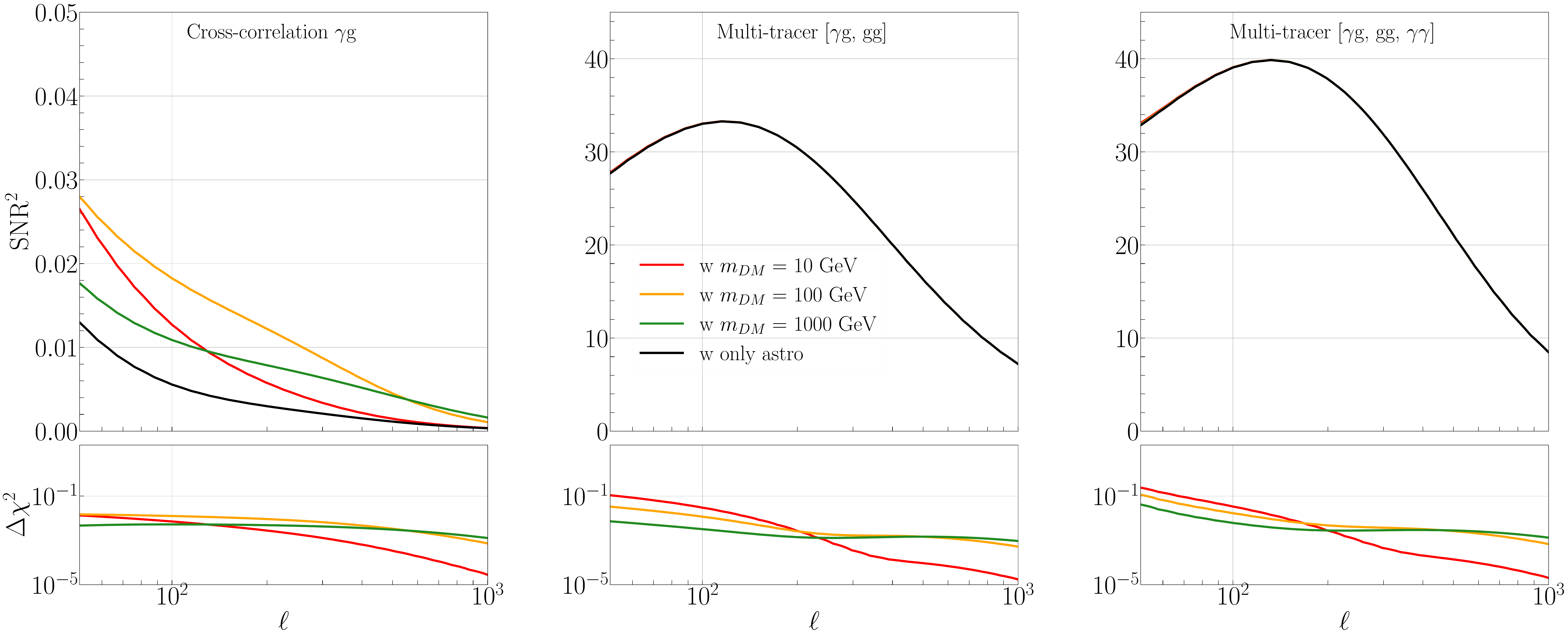}
    \caption{The same as in Fig. \ref{fig:SNR_noweight}, for the {\it DM+astro weighted} case.}
    \label{fig:SNR_weight_astroDM}
\end{figure}

\begin{figure}[t!]
    \centering
    \includegraphics[width=1\linewidth]{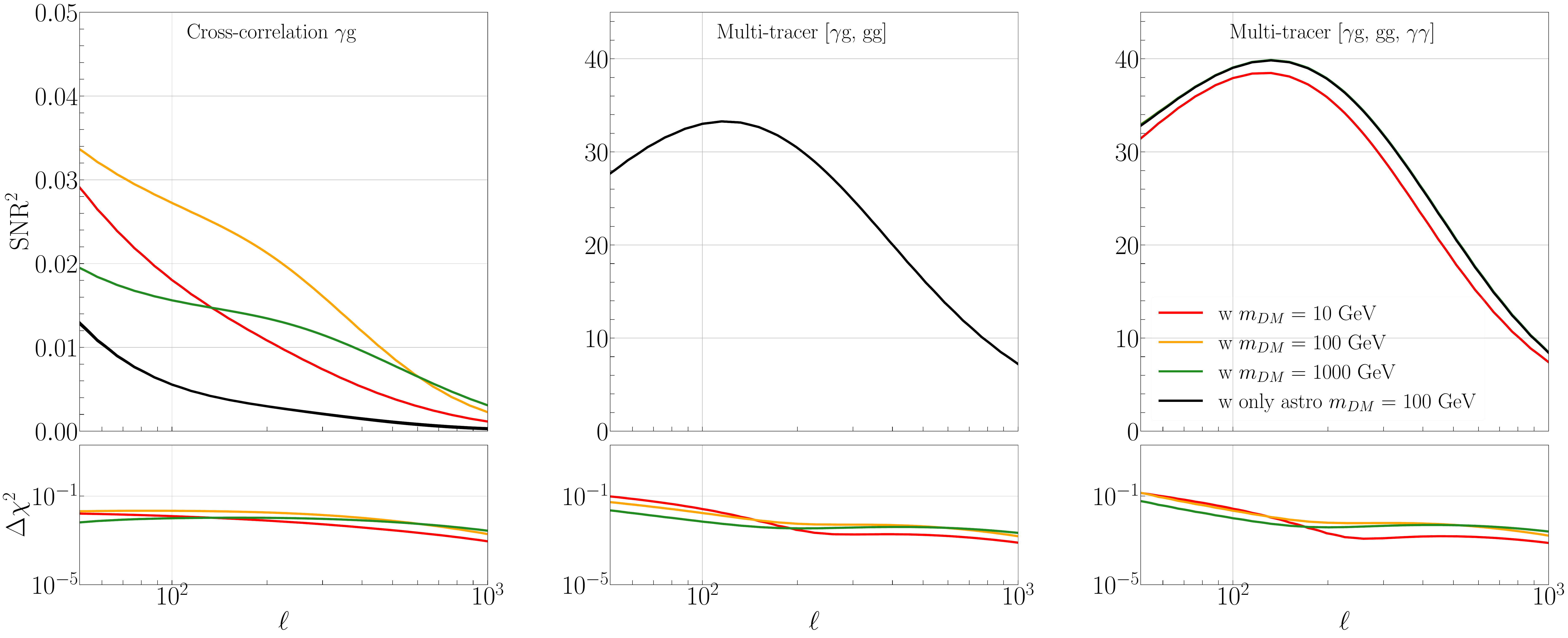}
    \caption{The same as in Fig. \ref{fig:SNR_noweight}, for the {\it DM-only weighted} case.}
    \label{fig:SNR_weight_DM}
\end{figure}

\begin{figure}[t!]
	\centering 
	\includegraphics[width=1
    \textwidth, angle=0]{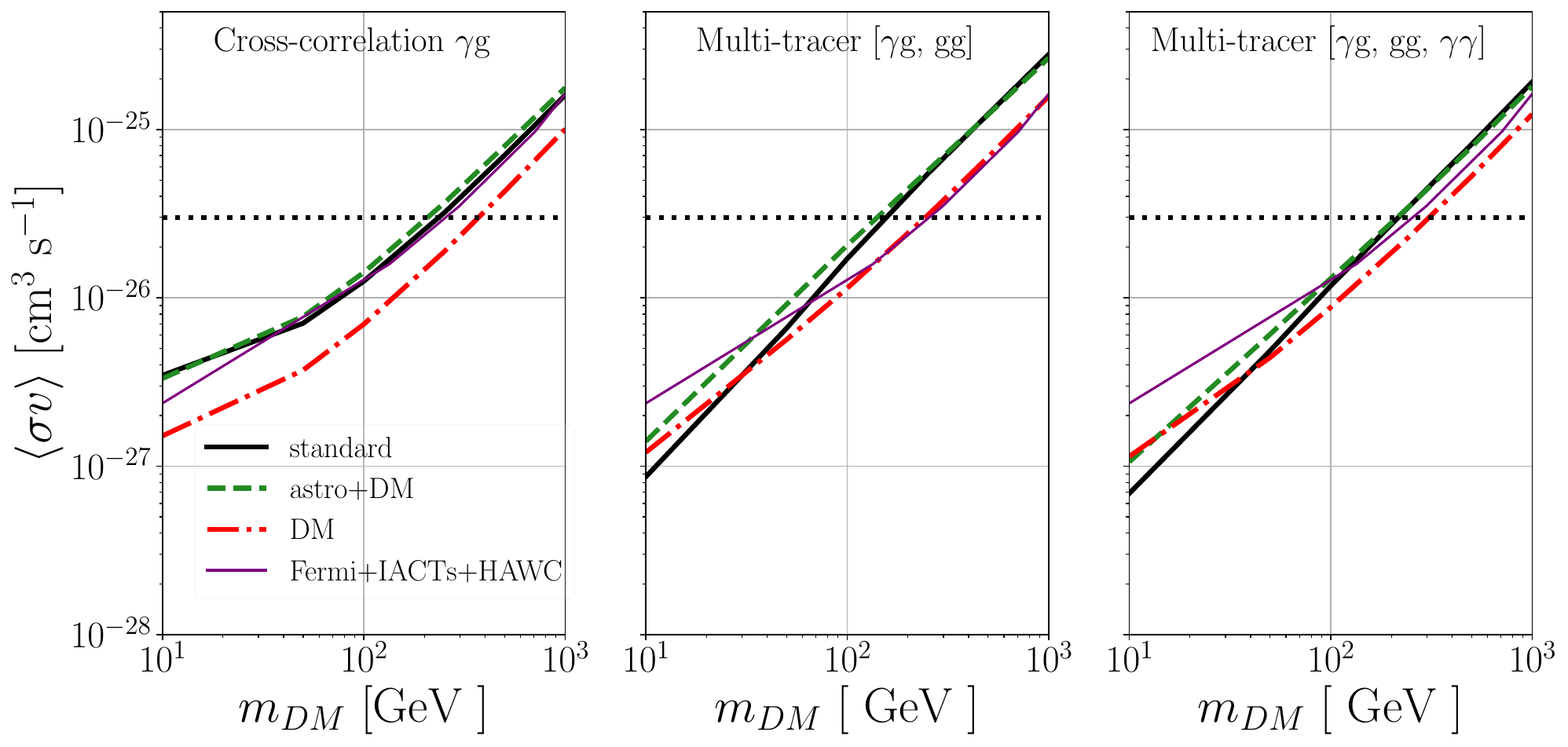}	
	\caption{Forecasts for the bounds on the dark matter annihilation cross section $\sv$ as a function of the DM mass $m_{\rm DM}$ under different analysis strategies and for the annihilation channel into $b \bar b$. The three panels refer (from left to right) to the adoption of the gamma-rays $\otimes$ galaxies cross-correlation ($\gamma g$), the multi-tracer that combines gamma-rays $\otimes$ galaxies cross-correlation with galaxies auto-correlation ($\gamma g$ + $gg$) and the multi-tracer that combines gamma-rays $\otimes$ galaxies cross-correlation, galaxies auto-correlation and gamma rays auto-correlation ($\gamma g$ + $gg$ + $\gamma\gamma$). In each panel, the (black) solid lines refers to the standard (unweighted) case, the (green) dashed lines refer to the case weighted {\it Astro+DM} case and the (red) dot-dashed lines refer to the weighted {\it DM-only} case. The thin solid line shows the current combined bound from searches towards dwarf spheroidal galaxies with Fermi-LAT, HAWC, H.E.S.S., MAGIC, and VERITAS \cite{Hess:2021cdp}, as recompiled in Ref. \cite{Cirelli:2024ssz}. The horizontal dotted line refers to the ``thermal'' annihilation cross section $\svth = 3\times 10^{-26}$ cm$^{-3}$ s$^{-1}$.}
	\label{fig:bounds}
\end{figure}

As a first result, we show in Figs. \ref{fig:SNR_noweight}, \ref {fig:SNR_weight_astroDM} and \ref{fig:SNR_weight_DM} the contributions arising from each multipole to the SNR$^2$ as defined in Eq. (\ref{eq:SNR_ell}) (upper panels) and to the $\Delta\chi^2$ defined in Eq. (\ref{eq:delta_chi_2}). The left panels of each figure refer to the gamma-rays $\otimes$ galaxies cross-correlation ($\gamma g$), the central panel to the multi-tracer technique that combines gamma-rays $\otimes$ galaxies cross-correlation with galaxies auto-correlation ($\gamma g$ + $gg$) and the right panel to the multi-tracer technique that adds gamma rays auto-correlation ($\gamma g$ + $gg$ + $\gamma\gamma$). In each upper panel, the black lines refer to gamma-ray emission from astrophysical sources only. The red, yellow and green lines refer to the cases in which a dark matter gamma-ray emission is added to the underlying astrophysical sources emission. The three lines refer to three representative dark matter mass values and annihilation cross sections: 10 GeV and $\sv = 0.05 \times \svth$ (red), 100 GeV and $\sv = 0.5 \times \svth$ (yellow), 1000 GeV and $\sv = 5 \times \svth$ (green), where $\svth = 3\times 10^{-26}$ cm$^{-3}$ s$^{-1}$ is the ``thermal'' value which corresponds to a dark matter particle relic abundance that matches the observed dark matter content of the Universe. These representative values for $\sv$ are chosen such that they fall close to the values of $\sv$ that will correspond to the predicted sensitivity at each of the three mass values, as it will be discussed in connection with Fig. \ref{fig:bounds}. Let us also remind that in Fig. \ref{fig:SNR_weight_DM} we have a different $\snr^2_{\rm{ astro}}$ for each DM mass, since in the {\it DM-only} case the filter is based on the window function of the DM gamma-ray emission, which is defined only for $E < m_{\rm DM}$ (``capped'' $\snr^2_{\rm{ astro}}$). The difference among the various ``capped'' $\snr^2_{\rm{ astro}}$ is nevertheless very small and therefore only one case is shown in the upper panels of the figure. In the lower panels, where the $\Delta\chi^2$ is shown, we have nevertheless used the actual value of the various ``capped'' $\snr^2_{\rm{ astro}}$ to determine the $\Delta\chi^2$.

We notice that the dominant contribution to the SNR$^2$ in the case of the $(\gamma g)$ cross-correlation comes from low multipoles, with a decreasing behaviour in multipole, with some small features for the case in which a gamma-ray DM emission is present, especially at higher masses. On the contrary, in the two multi-tracer cases, the SNR$^2$ has a pronounced dominant peak for multipoles around 100-200, mostly driven by the inclusion of the galaxy-galaxy autocorrelation observable. While the absolute value of the SNR$^2$ is largely increased in the case of multi-tracers, as a consequence of the high significance of the auto-correlation signal of galaxies and gamma-rays, the relevant quantity that allows us to infer bounds for the DM parameters is the $\Delta\chi^2$. Figs. \ref{fig:SNR_noweight}, \ref {fig:SNR_weight_astroDM} and \ref{fig:SNR_weight_DM} show that the multi-tracer technique can significantly increase the contribution to the $\Delta\chi^2$. The increase is mostly driven at the lower multipoles, where cosmic variance is more prominent and the inclusion of additional information from the multi-tracers allows to reduce the variance. On the contrary, at large multipoles, where the impact of the \Fermi\  PSF becomes increasingly relevant, thus boosting the variance and consequently depressing the SNR$^2$, the adoption of a multi-tracer technique is not especially effective in improving the statistical significance. 

Coming now to the impact of the filters, Figs. \ref{fig:SNR_noweight}, \ref {fig:SNR_weight_astroDM} and \ref{fig:SNR_weight_DM} show that they can indeed be relevant, although depending on the specific type of analysis. In the case of the cross-correlation $(\gamma g)$ alone, the {\it DM-only} filter can improve the SNR$^2$ over the whole multipole range (let us stress that, even though the single contribution to the $\Delta\chi^2$ from each multipole is small, the cumulative value of Eq. (\ref{eq:SNR2}) is the one that matters, and it is obtained by adding contributions over a wide multipole range, from 50 to 1000). Therefore, in the pure cross-correlation case, the results show that the adoption of the filtering technique allows us to improve the statistical significance, as compared to the standard method of analysis usually adopted, which does not rely on the use of a filter. In the case of the multi-tracer $(\gamma g + gg)$, the most prominent improvement occurs again for the {\it DM-only} filter and mostly for large DM masses, where the high-multipole tail of the $\Delta\chi^2$ can be increased for a wide interval of multipoles. The addition of the gamma-ray autocorrelation observables in the $(\gamma g + gg + \gamma\gamma)$ multi-tracer allows to further slightly improve the significance for a DM signal. In this case we add a potentially relevant amount of information from the correlations between different energy bins of the gamma-ray auto-correlation signal (see the expression of the covariance matrix and its dimensions, Eq. (\ref{eq:matrixC}) and (\ref{eq:matrixC_dim})) and this correlated information among different energy bins can actually help in separating contributions endowed with different spectral behaviours, like in the case of DM vs astrophysical sources emissions.

The forecasts on the bounds that can be achieved with the different strategies are shown in Fig. \ref{fig:bounds}. The three panels again refer, from left to right, to the ($\gamma g$), ($\gamma g$ + $gg$) and ($\gamma g$ + $gg$ + $\gamma\gamma$) cases. In each panel, the different lines refer to the three strategies outlined in Table \ref{tab:strategies}: the (black) solid lines refer to the {\it unweighted} case, the (green) dashed lines refer to the {\it Astro+DM} weight and the (red) dot-dashed lines refer to the {\it DM-only} weight. We observe that in the case of $(\gamma g)$ cross-correlation alone, the adoption of the filter {\it DM-only}, based on the predicted gamma-ray DM emission, would allow us to improve bounds by a factor of ranging from 2 to 2.5 over the whole mass range, as compared to the standard analysis where no filter is applied. At the same time, the adoption of a multi-tracer technique can be very effective, even without further adopting a filter, for low DM masses: the ($\gamma g$ + $gg$) multi-tracer allows to improve the bound for low DM masses by up to a factor of 4, and the ($\gamma g$ + $gg$ + $\gamma\gamma$) can enhance this improvement up to a factor of 5. For higher masses, ($\gamma g$), ($\gamma g$ + $gg$) and ($\gamma g$ + $gg$ + $\gamma\gamma$) all give comparable bounds, when the {\it DM-only} filter is applied. From our analysis, we instead find that the filter {\it DM+astro}, based on both the DM and astrophysical sources window functions, is never specifically effective for improving the DM bounds, despite the fact that it contains the different behaviours of the astrophysical source and DM window functions as a function of redshift and of energy, as shown in Fig. \ref{fig:UGRB_windows} and \ref{fig:UGRB_windowsE}.  

This can be related to the fact that here we are modeling our galaxy catalog window function on 2MRS, which covers a restricted redshift range. While 2MASS focuses on low redshift, which is where the DM gamma-ray signal is mostly prominent, contrary to the emission from unresolved astrophysical sources that instead peaks at higher redshift, nevertheless the restricted range does not allow to leverage on the different dependence of DM and astrophysical sources with redshift. Galaxy catalogs with a wider redshift range, still covering the low-redshift regime optimal for DM cross-correlation searches, would allow to add a further dimension to the analysis. This specific point is left for a future investigation.

In summary, from the whole analysis discussed above, we can devise that the best strategy to improve the reaching capabilities of the correlation technique for the search of a DM signal is the adoption of a weighting technique where the filter is modeled on the theoretical expectations of the gamma-ray DM emission window function, i.e. on the {\it DM-only} method, whose features are outlined in Table \ref{tab:strategies}. This method allows to improve the constraint on the DM annihilation cross section $\sv$ by about a factor ranging from 2 to 2.5 over the whole mass range, as compared to the standard correlation technique based on the gamma-rays $\otimes$ galaxies cross-correlation technique ($\gamma g$). Nevertheless, to further improve the bounds for DM masses below about 50 GeV, the full multi-tracer method ($\gamma g$ + $gg$ + $\gamma\gamma$), even without the adoption of a filter, is very effective, improving the bounds for light DM masses up to a factor of 5. The overall best strategy would therefore be the implementation and combination of both the filtering and the full multi-tracer methods, due to their complementarity discussed above.

\section{Conclusions}
\label{sec:conclusions}

In this paper, we have investigated two methodologies to improve the sensitivity of the cross-correlation between a gravitational tracer of dark matter, specifically a galaxy catalog, and a dark-matter particle-physics signal, namely the emission of a gamma-ray cosmological background produced by the annihilation of weakly interacting massive particles responsible for the dark matter. The case study has been focused on a low-redshift galaxy catalog, modeled on the 2MRS survey, and on the \Fermi \ gamma-ray data set, with a detector sensitivity forecast to 20 years of data taking.

From one side, we have studied the impact of adopting a Wiener filter on the measured redshift distribution of the galaxy catalog, based on the expected gamma-ray emission. This method requires to apply a specific redshift-dependent weighting scheme to the galaxy distribution, for which we have investigated different strategies. From the other side, we have investigated if the adoption of the full multi-tracer information available, i.e. the proper combination of the auto- and cross-correlations of the two observables (galaxy distribution and unresolved gamma-ray background map), can help in improving the sensitivity of the cross-correlation analysis.

We have found that the adoption of a weighting scheme based on the emission model of dark matter alone (not in combination with the astrophysical sources emission, i.e. {\it DM-only} technique) is the most effective in improving the sensitivity to the DM particle parameters when the cross-correlation observable is used as a single observable, i.e. not in combination with the full multi-tracers. This method allows to improve by a factor  ranging from 2 to 2.5 the constraint on the DM annihilation cross section $\sv$ over the whole mass range, as compared to the standard correlation technique based on the gamma-rays $\otimes$ galaxies cross-correlation technique. The adoption of the full multi-tracers information, i.e. the combined analysis of the galaxy $\otimes$ gamma-ray cross-correlation with the galaxy $\otimes$ galaxy auto-correlation and gamma-ray $\otimes$ gamma-ray autocorrelation, allows to improve the bounds for DM masses below about 50 GeV up to a factor of 5 as compared to the standard analysis (gamma-ray $\otimes$ galaxies), even without the adoption of the filter. The {\it DM-only} filter becomes again instrumental in improving the bounds for heavier dark matter masses: nevertheless, in this case, the sensitivity reach is similar to what can be obtained by the adoption of the filter with the gamma-ray $\otimes$ galaxies cross-correlation. Concerning the robustness of the two proposed techniques, there are two aspects to be considered, as we are looking at the combined effect of optimal weights and a multi-tracer approach. Care for the latter has to be taken when considering auto-correlation spectra. Indeed, cross-correlation spectra are known to be very robust to systematic errors, as additive systematics in any of both fields will drop in the cross-correlation (assuming systematics do not correlate, which is expected to be largely true in our case since we are considering cross-correlation of fields which are measured in a totally independent way) . 
On the other hand, auto-correlation spectra need to be modeled carefully, as any residual systematics will propagate into the multi-tracer result \cite{Cross_Alimi_Valls_Blanchard,Theroleofcross_Blanchard}. This notwithstanding, the observables of our analysis are well-known astrophysical and cosmological observables, and their separate analyses (auto-correlation spectra) have been carried out and confronted with theoretical models in an extensive set of analyses, ensuring a high degree of reliability of the modeling employed \citep{Fermi-LAT:2018udj, Ando_2MRS_2018}. 
Coming to the Wiener filters, the main assumption is the knowledge of the gamma-ray kernel, by which we weight the galaxy one so that the two kernels can match \cite{Fisher_Lahav_Wiener,Lahav_Fisher_Wiener,Zaroubi_Hoffman_Fisher_Wiener,Alonso_Wiener_GW,}. If our knowledge is unsatisfactory, the real kernels of the gamma-ray emission will not overlap with that of the weighted galaxies. However, the outcome of this would result into a sub-optimal weighting and, as a consequence, the SNR will not increase and the improper modeling will not be selected by the technique. Iteratively, one can try and optimise the matching - effectively following a matched-filter approach, which is what we are proposing here. 

In conclusion, we have shown that the two techniques discussed in this paper, namely the adoption of a weighting scheme based on a Wiener filter formalism and the adoption of the full set of correlations relative to our observables (multi-tracer) can indeed improve the capabilities of cross-correlations between the UGRB and galaxy catalogs in setting bounds on particle dark matter. This methodology can be directly extended to comprise different gravitational tracers of dark matter, like galaxy-cluster catalogs or cosmic shear, and different electromagnetic bands for the emission from dark matter, like X-rays or radio emission, therefore being relevant also for dark matter candidates different from the WIMP case discussed in this paper, like axions and axion-like-particles. In addition, the Wiener filter technique and the multitracer analysis applied to the cross-correlation technique could be successfully adopted to improve the sensitivity to the searches for a dark matter signal in non-electromagnetic channels, like neutrinos and charged cosmic-ray emissions \cite{urban2023modelling}, and could be applied also to the study of the unresolved population of astrophysical sources in the whole multi-messenger and multi-wavelength scenario: in this case, galaxy catalogs peaked at redshift higher than 2MRS would be more suitable, since the emissions from unresolved astrophysical sources tend to be dominated at larger redshifts than in the case of dark matter.

\appendix
\section{Dark matter}
\label{app:dark_matter}

As discussed in the main text, we model the dark matter distribution in the Universe in the framework of the halo model \citep{cooray2002,Asgari_halo_model_2023}. A halo is defined as a spherical dark matter overdensity with mass $M$ and size defined by its virial radius $ r_{\rm{vir}} = \left[3 M /4 \pi \rho_{\rm c}(z)  \delta_{c}(z) \right]^\frac{1}{3}$, where 
$\rho_{\rm c}(z)$ is the Universe critical density at redhift $z$ and $\delta_c(z)$ is the overdensity required to form a halo, for which we adopt $\delta_{c}(z) = 18 \pi^2 - 82\,\Omega_{\Lambda}(z) - 39\,\Omega^2_{\Lambda}(z)$ \citep{Bryan_Norman_1998}, where $\Omega_{\Lambda}(z)$ is the dark energy density parameter. 

The halo density profile $\rho(r)$ is modeled as a NFW profile \citep{navarro1996,NFW_1997}, normalized to the halo mass $M$ inside its virial radius. The scale radius $r_s$ of the NFW profile is obtained from the virial radius through the halo-mass-dependent concentration parameter $c_{\rm vir} = r_{\rm vir}/r_s$, for which we adopt the halo concentration model of Ref. \citep{Correa_2015}. Since Ref. \citep{Correa_2015} defined halos by means of an overdensity of 200 with respect to the critical density, we relate the $c_{200}$ concentration parameter of Ref. \citep{Correa_2015} to $c_{\rm vir}$ as: $ c_{\rm{vir}} = b + a \ c_{\rm{200}} \label{eq:cv_c_200}$, with $a = -1.119 \log_{10}(\delta_{\rm{c}}) + 3.537$, $b = -0.967 \log_{10}(\delta_{\rm{c}}) + 2.181$ \citep{Coe2010}.

For annihilating dark matter, the emitted gamma ray flux is proportional to the square of the dark matter density. Therefore, the window function Eq. (\ref{eqn:Wann}) can be cast in terms of the so-called clumping factor:\begin{equation}
    \Delta^2(z) \equiv \frac{\langle \rho^2(z) \rangle}{\langle \rho(z) \rangle^2}, \label{eq:clumping}
\end{equation}
where
\begin{equation}
\langle \rho^2(z) \rangle = \int \de M \frac{dn}{d M}(z)\int \de ^3 \bm{r} \ \rho^2(\bm{r}, M, z)\;,
\label{eq:gave}
\end{equation}
and 
$\langle \rho(z) \rangle$ is the corresponding integral for the dark matter density. In Eq. (\ref{eq:gave}), $dn/dM$ is the halo mass function, for which we adopt Ref. \citep{sheth_tormen_1999}. Since dark matter halos can contain substructures, we model the presence of subhalos by modifying the internal halo mass density squared as: $\rho^2(\boldsymbol{r},z,M) \rightarrow \rho^2(\boldsymbol{r},z,M)[1 + B(M,z)]$ and we adopt the boost function $B(M,z)$ of Ref. \citep{Moline_Sanchez_Conde_2017}:
\begin{equation}
    \log_{10} B(M,z) = \sum _{i = 0} ^5 b_i \left [ \log_{10} \left ( \frac{M(z)}{M_{\odot}} \right ) \right],
\end{equation}
with $b_i  = \left( - 0.186,\; 0.144,\; - 8.8 \cdot 10^{-3},\; 1.13 \cdot 10^{-3},\; - 3.7 \cdot 10^{-5},\; -2.0 \cdot 10^{-7} \right)$, where the values in parenthesis run over $i=0, ..., 5$. In all our analyses, we consider dark matter halos with masses from 
$10^{-6} M_\odot$ to $10^{18} M_\odot$.

\section{Astrophysical sources}
\label{app:astro_sources}

As astrophysical gamma-ray sources, we consider four classes: blazars (BLA), misaligned active galactic nuclei (mAGN), flat spectrum radio quasars (FSRQ) and star forming galaxies (SFG). 
The window function of the gamma-ray emission from astrophysical sources is outlined in Eq. (\ref{eq:wastro}) and repeated here for convenience:
\begin{equation}
     W_{\rm astro}(E, z) =  \frac{dW_{\rm astro}}{dEd\chi}= \left(\frac{d_L(z)}{1+z} \right)^2   \int_{\mathcal{L}_{\rm min}}^{\mathcal{L}_{\rm max}(z)} d \mathcal{L} \, \frac{d F}{d E}(E,\mathcal{L},z) \, \phi(\mathcal{L},z) e^{-\tau(E,z)}, \label{eq:wastroA}
\end{equation}
where the quantities that refer to the gamma-ray sources are: the source luminosity $\mathcal{L}$, the gamma-ray luminosity function $\phi(\mathcal{L},z)$ and the spectral energy distribution (SED) $dF/dE$.

\begin{table*}[t!]
\centering
\begin{tabular}{|l|ccc|}
\hline
Source class & $\Gamma$ & $\mathcal{L}_{\rm min}\,[\mathrm{erg\,s^{-1}}]$ & $\mathcal{L}_{\rm max}\,[\mathrm{erg\,s^{-1}}]$ \\ 
\hline
BL Lacs & 2.11 & $7 \times 10^{43}$ & $10^{52}$ \\ 
mAGN & 2.37 & $10^{40}$ & $10^{50}$ \\
FSRQ & 2.44 & $10^{44}$ & $10^{52}$ \\
SFG & 2.7 & $10^{37}$ & $10^{42}$ \\
\hline
\end{tabular}
\caption{Spectral index $\Gamma$, minimal luminosity $\mathcal{L}_{\rm min}$ and maximal luminosity $\mathcal{L}_{\rm max}$ four the four astrophysical contributors to the unresolved gamma ray background considered in the paper.}
\label{tab:astroparam}
\end{table*}

By convention, the luminosity $\cal L$ of the source is defined as integrated over the energy interval (0.1, 100) GeV:
\be
{\cal L} = \int_{0.1\, {\rm GeV}}^{100\, {\rm GeV}} \de E_r \, \frac{\de F}{\de E_r}
\ee
where $E_r$ is the photon energy in the rest frame of the source, related to the observed energy by $E_r = (1+z) E$, and $\de F/\de E$ is the spectral energy distribution of the sources, which we model here as a power law with index $\Gamma$:
\begin{equation}
\frac{\de F}{\de E} = A({\cal L},z) \; E^{-\Gamma}
\label{eq:SED}
\end{equation}
The values of $\Gamma$ and $E_{\rm cut}$ for the four classes of astrophysical sources adopted in our model are listed in Table \ref{tab:astroparam}, together with the minimal and maximal  luminosities of the source classes ${\cal L}_{\rm min}$ and ${\cal L}_{\rm max}$. The amplitude $A({\cal L},z)$, instead, can be related to the luminosity ${\cal L}$ of the source by observing that \cite{Hogg:1999ad,Pinetti:2025qpi}:
\be
{\cal L} = \frac{4\pi d_L^2(z)}{(1+z)} \int_{0.1\, {\rm GeV}}^{100\, {\rm GeV}} \de E_r \, E_r \, \frac{\de F}{\de E_r} = 4\,\pi\,d_L^2(z) \int_{0.1\, {\rm GeV}/(1+z)}^{100\, {\rm GeV}/(1+z)} \de E \, E \, \frac{\de F}{\de E}
\label{eq:luminosity}
\ee
from which we obtain:
\be
A({\cal L},z) = \frac{{\cal L}}{4\,\pi\,d_L(z)^2} 
\left[ 
\int_{0.1\, {\rm GeV}/(1+z)}^{100\, {\rm GeV}/(1+z)} 
\de E \,  E^{(-\Gamma + 1)} \right]^{-1}
\ee
Therefore, the SED reads:
\begin{equation}
    \frac{d F}{d E}(E,\;\mathcal{L},\;z)=\frac{\mathcal{L}}{4\pi d_L^2(z)} \left\lbrace \frac{1}{2-\Gamma}  \left[ \left(    \frac{100}{1+z}\right)^{2-\Gamma}-\left(\frac{0.1}{1+z}\right)^{2-\Gamma}\right]\right\rbrace^{-1}\left(\frac{E}{\text{GeV}}\right)^{-\Gamma} 
    \label{eq:SED:Pinetti} .
\end{equation}

The maximal luminosity $L_{\rm max}(z)$ in Eq. (\ref{eq:wastroA}),  instead, depends on the threshold of detectability of resolved sources by the detector (since we are dealing with the unresolved gamma-ray background) as:
\begin{equation}
{\cal L}_{\rm max}(z) = \min({\cal L}_{\rm max},{\cal L}_{\rm sens}(z))
\end{equation}
where ${\cal L}_{\rm max}$ is given in Table \ref{tab:astroparam} for each source class and ${\cal L}_{\rm sens}(z)$ is obtained as:
\be
{\cal L}_{\rm sens}(z) = 4\,\pi\,d_L^2(z) \; F_{\rm sens}^{\rm LAT} \; 
\frac{
\int_{0.1\, {\rm GeV}/(1+z)}^{100\, {\rm GeV}/(1+z)} \de E \, E^{-\Gamma+1}}
{\int_{1\, {\rm GeV}}^{100\, {\rm GeV}} \, \de E \, E^{-\Gamma} \, \exp[-\tau(E,z)]}
\label{eq:threshold}
\ee
where $F_{\rm sens}^{\rm LAT}$ is the \Fermi \ flux sensitivity in the energy interval (1,100) GeV. We assume $F_{\rm sens}^{\rm LAT} = 10^{-10} {\rm cm}^{-2} {\rm s}^{-1}$, a slightly better improvement than the current sensitivity \cite{Manconi:2019ynl}. While the flux sensitivity can have a dependence with the source spectral index, we adopted the common assumption of a flat value for $F_{\rm sens}^{\rm LAT}$. For other approaches, see e.g. Ref. \cite{Manconi:2019ynl}. 

\subsection{Mass-to-luminosity relation}
\label{sec:Mass_luminosity_relation} 
The halo model for astrophysical objects requires a conversion between the luminosity of the source $\cal L$ and the mass $M$ of the halo that hosts the same source. In our modeling we have followed Ref. \citep{Pinetti_thesis_2022}, to which we refer for additional details.

\paragraph{BL Lac and Flat Spectrum Radio Quasars.} For BL Lacs and FSRQ, we adopt a common expression for the mass-to-luminosity ratio \citep{Pinetti_thesis_2022}:
\begin{equation}
M(\lum)=10^{13} M_{\odot}\left(\frac{M_{\star}}{10^{8.8}(1+z)^{1.4}}\right)^{0.645}
\label{M_to_L_BL_RQ}
\end{equation}
with
\begin{equation}
M_{\star}=10^{9}\left(\frac{\lum}{10^{48} \mathrm{erg} / \mathrm{s}}\right)^{0.36}  \label{eq:Mstar_to_L_Bl_RQ}    \end{equation}

\paragraph{Misaligned active galactic nuclei.}For mAGNS, we adopt the following expression \citep{Pinetti_thesis_2022}:

\begin{equation}
M(\lum)=10^{13} M_{\odot}\left(\frac{M_{\star}}{10^{8.8}(1+z)^{1.4}}\right)^{0.645}
\label{eq:M_to_L_mAGN}
\end{equation}
with
\begin{equation}
M_{\star}=4.6 \cdot 10^{9}\left(\frac{\lum}{10^{48} \,\mathrm{erg} / \mathrm{s}}\right)^{0.16}
\label{eq:Mstar_to_L_mAGN}    
\end{equation}

\paragraph{Star forming galaxies.} For SFG we adopt the following relation \citep{Pinetti_thesis_2022}:

\begin{equation}
M(\lum)=\frac{10^{12} M_{\odot}}{(1+z)^{1.61}}\left(\frac{\lum}{6.8 \cdot 10^{39} \,\mathrm{erg} / \mathrm{s}}\right)^{0.92}
\label{eq:M_to_L_SFG}
\end{equation}

\subsection{Gamma-ray Luminosity Functions}

The gamma-ray luminosity function (GLF) $\phi(\mathcal{L},z)$ defines the number density of astrophysical objects per unit luminosity, unit volume and unit spectral index:
\begin{equation}
\phi({\cal L}, z, \Gamma) = \frac{\de N}{\de {\cal L} \,\de V \,\de \Gamma}
\end{equation}
Again following Ref. \citep{Pinetti_thesis_2022}, we list below the GLF of each source class we adopt in our analysis. 

\paragraph{BL Lac and Flat Spectrum Radio Quasars.}

We adopt the luminosity-dependent density evolution (LDDE) model of Ref. \citep{ajello2014} for BL Lacs and Ref. \citep{ajello2012} for FSRQ:
\begin{equation}
\phi_{\gamma}(\mathcal{L}, z, \Gamma)=\phi(\lum, \Gamma) \times e(z, \mathcal{L})
\label{eq:BL:phi_gamma_z}
\end{equation}
At redshift $z=0$, the GLF is parametrised as a broken power law in luminosity and it follows a Gaussian distribution in photon spectral index:

\begin{equation}
\phi(\mathcal{L}, \Gamma')=\frac{A}{\log (10)}\left(\frac{\mathcal{L}}{\rm{erg~s}^{-1}}\right)^{-1}\left[\left(\frac{\mathcal{L}}{\mathcal{L}_{\star}}\right)^{\gamma_{1}}+\left(\frac{\lum}{\mathcal{L}_{\star}}\right)^{\gamma_{2}}\right]^{-1} \exp \left[-\frac{(\Gamma'-\mu(\mathcal{L}))^{2}}{2 \sigma^{2}}\right]
\label{eq:BL_phi_L}
\end{equation}
where the mean spectral index $\mu$ has a weak (logarithmic) dependence on $\mathcal{L}$:
\begin{equation}
\mu(\mathcal{L})=\mu^{\star}+\beta_1\left[\log_{10} \left(\frac{\mathcal{L}}{\rm{erg~s}^{-1}}\right)-46\right]
\label{eq:mu_L}
\end{equation}

\begin{table*}[t!]
\centering
\begin{tabular}{|l|cccccccccc|}
\hline
 & $A$ & $\lum_{\star}\left[\mathrm{erg} \ \mathrm{s}^{-1}\right]$ & $\gamma_{1}$ & $\gamma_{2}$ & $p_{1}$ & $p_{2}$ & $z_{\star}$ & $\beta_1$ & $\beta_2$ & $\sigma$  \\
\hline
BL Lacs & $9.20 \cdot 10^{-11}$ & $2.43 \cdot 10^{48}$ & 1.12 & 3.71 & 4.50 & -12.88 & 1.67 & $0.0604$ & $0.0446$ & 0.26   \\
FSRQ & $3.06 \cdot 10^{-9}$ & $0.84 \cdot 10^{48}$ & 0.21 & 1.58 & 7.35 & -6.51 & 1.47 & 0. & 0.21 & 0.18  \\
\hline
\end{tabular}
\caption{Fiducial values for the model parameters of the gamma-ray luminosity functions of BL Lacertae and Flat-Spectrum Radio Quasars. The parameter $A$ is in units of $\mathrm{Mpc}^{-3} {\rm dex}^{-1}$. The $\lum_{\star}$ is in units of $\mathrm{erg} \ \mathrm{s}^{-1}$.}
\label{tab:BL_FSRQ_pars}
\end{table*}
The redshift dependence is encoded in the function $e(z, \mathcal{L})$ \citep{Pinetti_thesis_2022}: 
\begin{equation}
e(z, {\cal L})=\left[\left(\frac{1+z}{1+z_{c}(\lum)}\right)^{-p_{1}}+\left(\frac{1+z}{1+z_{c}(\lum)}\right)^{-p_{2}}\right]^{-1}
\label{eq:phi_BLRQ_z}
\end{equation}
where $z_{c}=z_{\star}\left(\lum / 10^{48} \,\mathrm{erg} / \mathrm{s}\right)^{\beta_2}$. BL Lacs and FSRQs exhibit the same functional form for the GLF, but they differ for the values of the parameters, which are displayed in Tab. \ref{tab:BL_FSRQ_pars}. 

In our analysis we have adopted this model with a slight modification. Instead of considering a gaussian distribution for the spectral indices as in Eq. (\ref{eq:BL_phi_L}), we have assumed a value for the spectral index of each source class fixed at its best fit value, without luminosity dependence, i.e. we have adopted the spectral indices of Tab. \ref{tab:astroparam}. This corresponds to the substitution in Eq. (\ref{eq:BL_phi_L}):
\begin{equation}
   \exp \left[-\frac{(\Gamma'-\mu(\mathcal{L}))^{2}}{2 \sigma^{2}}\right] \rightarrow \sqrt{2\pi}\sigma\,\delta_D(\Gamma' - \Gamma) 
   \label{eq:gaussdelta}
\end{equation}
with the index $\Gamma$ of Tab. \ref{tab:BL_FSRQ_pars} and the $\sqrt{2\pi}\sigma\,$ pre-factor to preserve the normalization.
Let us also comment that we have slightly rescaled the normalizations of BL Lac and FSRQ in order to fit the auto-correlation of UGRB as measured by \Fermi\ \cite{ackermann2018}. The rescaling factors are 0.494 for BL Lacs and 0.714 for FSRQ. Notice that alternatively, this rescaling is equivalent to set to 1 the exponential function of Eq. (\ref{eq:gaussdelta}) and rescale down the GLF of both BL Lacs and FSRQ by a common factor of 0.322. Finally, we highlight that the model is also well compatible with the measure intensity of the UGRB \cite{ackermann2015,Ammazzalorso:2018evf}.

\paragraph{Misaligned active galactic nuclei}

The GLF of mAGN can be derived from of the radio luminosity functions (RLF) of the same class of sources, whose knowledge is considered to be more solid than the one of the GLF. The two quantities are connected by the relation:
\begin{equation}
\phi_{\gamma}(\lum, z)=\frac{k}{\log (10) \mathcal{L}}\, \rho_{r}\left(\lum_{r}, z\right) \,\frac{\mathrm{d} \log_{10} \lum_{r}}{\mathrm{~d} \log_{10} {\cal L}},
\label{eq:phi_gamma_mAGN}
\end{equation}
where ${\cal L}_{r}$ is the radio luminosity, the constant $k$ is tuned to reproduce the numbers of mAGN observed by the $\gamma$-ray detector and $\rho_{r}\left(\lum_{r}, z\right)$ is the RLF, that, defined as the number of radio sources per unit of co-moving volume and per base-10 logarithmic unit of luminosity (dex), can be expressed as \citep{willot2001,Pinetti_thesis_2022}:
\begin{equation}
    \rho_{r}\left(\lum_{r}, z\right)=\rho_{l}\left(\lum_{r}, z\right)+\rho_{h}\left(\lum_{r}, z\right),
\label{eq:rho_r}
\end{equation}
where
\begin{eqnarray}
&\rho_{l}=\rho_{l \star}\left(\frac{\lum_{r}}{\lum_{l \star}}\right)^{-\beta_{l}} \exp \left(-\frac{\lum_{r}}{\lum_{l \star}}\right)(1+z)^{k_{l}} &\quad\text{for} \quad z<z_{l \star} \\ 
&\rho_{l}=\rho_{l \star}\left(\frac{\lum_{r}}{\lum_{l \star}}\right)^{-\beta_{l}} \exp \left(-\frac{\lum_{r}}{\lum_{l \star}}\right)\left(1+z_{l \star}\right)^{k_{l}} &\quad \text {for} \quad z \geq z_{l \star}
\label{eq:rho_ls}
\end{eqnarray}
and
\begin{equation}
\rho_{h}=\rho_{h \star}\left(\frac{\lum_{r}}{\lum_{h \star}}\right)^{-\beta_{h}} \exp \left(-\frac{\lum_{h \star}}{{\cal L}}\right) \,f_{h}(z)
\label{eq:rho_h}
\end{equation}
The function $f_{h}$ is defined as: 
\begin{equation}
f_{h}(z)=\exp \left\{-\frac{1}{2}\left(\frac{z-z_{h \star}}{z_{h 0}}\right)^{2}\right\}
\label{eq:f_h}
\end{equation}
The parameters of the previous equations are: 
$\rho_{l \star}=10^{-7.523} \,\mathrm{Mpc}^{-3}$,  
$\beta_{l}=0.586$, 
$\lum_{l \star}=10^{26.48} \,\mathrm{W} / \mathrm{(Hz \ sr)}$, 
$k_{l}=3.48$,  
$z_{l \star}=0.710$,  
$\rho_{h \star}=10^{-6.757} \mathrm{Mpc}^{-3}$,  
$\beta_{h}=2.42$,  
$z_{h \star}=2.03$ and
$\lum_{h \star}=10^{27.39} \,\mathrm{W} / \mathrm{(Hz \ sr)}$.
For $z<z_{h \star}$ we use $z_{h 0}=0.568$, while for $z \geq z_{h \star}$ we adopt $z_{h 0}=0.956$. Notice that, for consistency with the units of $\lum_{l \star}$ and of $\lum_{h \star}$, the luminosities in $\rho_{r, \rm{tot}}$ are expressed in units of $\rm{W \ Hz^{-1} \ sr^{-1}}$, that is, the luminosities in erg $\rm{s^{-1}}$ need to be first converted into W, then rescaled to the reference frequency value, 151 MHz, and divided by $4 \pi$.

Ref. \citep{Inoue:2011bm} and \citep{Dimauro2014} derived the correlation between the core radio and the $\gamma$-ray luminosities, with some updates to the formalism, adapting the formulae to the improved parameters values provided by Planck, whilst Ref. \citep{Lara:2004ee} found that the relations between the core and total luminosities read:
\begin{align}
\log_{10} \lum & =2+1.008 \log_{10} \lum_{r, \text { core }}^{5 \rm{GHz}} \label{eq:L_gamma_L_core}\\
\log_{10} \lum_{r, \text { core }}^{5 \mathrm{GHz}} & =4.2+0.77 \log_{10} \lum_{r, \text { tot }}^{1.4 \mathrm{GHz}} \label{eq:L_core_to_L_tot} 
,
\end{align}
where the relation between core and total RLF are related as:
\begin{equation}
\rho_{r, {\rm core}}(\lum_{r, \rm core}, z)=
\rho_{r, {\rm tot}}(\lum_{r, \rm tot}, z)
\frac{\mathrm{d} \log_{10} \lum_{r, \rm tot}}{\mathrm{~d} \log_{10} {{\cal L}_{r,\rm core}}}.
\label{eq:phi_gamma_mAGN}
\end{equation}
In Eq. (\ref{eq:L_gamma_L_core}), the luminosities are expressed in units of $\rm{erg \ s^{-1}}$, whilst in Eq. (\ref{eq:L_core_to_L_tot}) they are expressed in units of $\rm{W / (Hz )}$, 
Again, $151 \;\mathrm{MHz}$ is the reference frequency: hence after having solved Eqs. (\ref{eq:L_gamma_L_core}) and (\ref{eq:L_core_to_L_tot})
for $ \lum_{r, \text { tot }}^{1.4 GHz}$ as a function of $ \lum_{r, \text { core }}^{5 GHz}$ and of $\lum$, such value has to be expressed in terms of the luminosity at 151 MHz as \citep{Inoue:2011bm}:
\begin{equation}
\lum_{r} = \lum_{\mathrm{r}}^{\rm{1.4 GHz}} \left (\frac{\nu}{1.4 \rm{GHz}} \right) ^{-\alpha_{r}}, \label{eq:L_nu}
\end{equation}
where $\alpha_{r}=0.80$ for the total radio emission. The original relation used in Ref. \citep{willot2001} needs to be updated in agreement with the reference Planck 2015 cosmology, by relating the comoving volume element adopted in Ref. \citep{willot2001}:

\begin{equation}
\frac{\mathrm{d}^{2} V_{W}}{\mathrm{~d} z \mathrm{~d} \Omega}=\frac{c^{3} z^{2}(2+z)^{2}}{4 H_{0, W}^{3}(1+z)^{3}}
\label{eq:volume_element_w}
\end{equation}
where $H_{0, W}=50\, \mathrm{km} \,\mathrm{s}^{-1} \,\mathrm{Mpc}^{-1}$, and the comoving volume in the standard $\Lambda \mathrm{CDM}$ cosmology:
\begin{equation}
\frac{\mathrm{d}^{2} V}{\mathrm{~d} z \mathrm{~d} \Omega}=\frac{c d_{L}^{2}(z)}{H_{0}(1+z)^{2} \sqrt{\left(1-\Omega_{\Lambda}-\Omega_{m}\right)(1+z)^{2}+(1+z)^{3} \Omega_{m}+\Omega_{\Lambda}}}
\label{eq:volume_element}
\end{equation}
Thus,  defining a conversion factor:

\begin{equation}
\eta=\frac{\mathrm{d}^{2} V_{W} / \mathrm{d} z \mathrm{~d} \Omega}{\mathrm{d}^{2} V / \mathrm{d} z \mathrm{~d} \Omega},
\label{eq:eta}
\end{equation}
the GLF can be therefore be expressed as: 

\begin{equation}
\phi_{\gamma}(\lum, z)=\frac{k\, \eta}{(1+z)^{2-\Gamma}} \frac{1}{\ln (10) \,\lum_{\mathrm{tot}}^{151 \mathrm{MHz}}} \frac{\mathrm{d} \lum_{\mathrm{tot}}^{151 \mathrm{MHz}}}{\mathrm{d} L} \,\rho_{r, \rm{tot}}\left(\lum_{\mathrm{tot}}^{151 \mathrm{MHz}}(\lum)\right), 
\label{eq:phi_gamma_mAGN}
\end{equation}
where $k=3.05$. For the spectral index we adopt $\Gamma=2.37$. The factor $(1+z)^{2-\Gamma}$ is the so-called K-correction, that takes into account the redshift variation between observed and emitted energies.

\paragraph{Star forming galaxies.}

For SFG, we adopt a similar technique, by linking the GLF to the infrared luminosity function, for which we adopt the model of Ref. \citep{gruppioni2013}. This model identifies three separate and independently evolving sub-classes of SFG: quiescent spiral galaxies, starburst galaxies and SFG hosting a concealed or low-luminosity AGN  (SF-AGN). The total IR luminosity function is the sum of these three contributions:

\begin{equation}
\hat{\phi}_{\mathrm{IR}}=\hat{\phi}_{\text {spiral }}+\hat{\phi}_{\text {starburst }}+\hat{\phi}_{\text {SF-AGN }} .
\label{eq:phi_hat_ir_SFG}
\end{equation}

For each component we can model the infrared luminosity function as:

\begin{equation}
\hat{\phi}_{i}=\hat{\phi}_{0, i}\left(\frac{\lum_{\mathrm{IR}}}{\lum_{0, i}}\right)^{1-\gamma_{i}} \exp \left(-\frac{1}{2 \sigma_{i}^{2}} \right)\log _{10}^{2}\left(1+\frac{\lum_{\mathrm{IR}}}{\lum_{0, i}}\right)\
\label{eq:phi_i_ir_SFG}
\end{equation}
where $i=$ spiral, starburst and SF-AGN.  
$\lum_{0, i}$ for spirals is defined in the following way: 
\begin{equation}
\lum_{0, \rm spiral}= 
\left\{
\begin{aligned}
&\lum_{\star, sp}\left(\frac{1+z}{1.15}\right)^{k_{L sp}} & \text { for } z \leq 1.1 \\ 
&\lum_{\star, sp}\left(\frac{2.1}{1.15}\right)^{k_{L sp}} & \text { for } z \geq 1.1 \\
\end{aligned}
\right.
\label{eq:L_0_i_spirals}
\end{equation}
while for starburst and SF-AGN we have: 
\begin{equation}
\lum_{0, i} =  \lum_{\star, i}  \left(  1 + z   \right) ^ {k_{L i}}.
\label{eq:L_0_starburst_SFAGN}
\end{equation}
The quantity $\hat{\phi}_{0, i}$ for spirals reads:
\begin{equation}
\hat{\phi}_{0, \text {spiral}}= 
\left\{
\begin{aligned}
&\phi_{\star, \mathrm{sp}}\left(\frac{1+z}{1.15}\right)^{k_{R 1, \mathrm{sp}}} && \text { for } z \leq 0.53 \\ 
&\phi_{\star, \mathrm{sp}}\left(\frac{1.53}{1.15}\right)^{k_{R 1, \mathrm{sp}}}\left(\frac{1+z}{1.53}\right)^{k_{R 2, \mathrm{sp}}} && \text { for } 0.53 < z \leq 1.7 \\
& 0  && \text { for } z > 1.7
\end{aligned}
\right.
\label{eq:phi_0_spiral}
\end{equation}
while for starburst and SF-AGN we have:
\begin{equation}
\hat{\phi}_{0, j}= 
\left\{
\begin{aligned}
&\phi_{\star, j}\left(\frac{1+z}{1.15}\right)^{k_{R 1, j}} & \text { for } z \leq 1.1 \\ 
&\phi_{\star, j}\left(\frac{2.1}{1.15}\right)^{k_{R 1, j}}\left(\frac{1+z}{2.1}\right)^{k_{R 2, j}} & \text { for } z>1.1
\end{aligned}
\right.
\label{eq:phi_0_j}
\end{equation}
Since 2MRS covers redhisfts below 0.1, we adopt the model parameters of Ref. \citep{gruppioni2013} for the first redshift bin ($0 \leq z \leq 0.3$ ) and list them in Tab. \ref{tab:params_agn}.

\begin{table*}[t!]
\centering
\begin{tabular}{|l|ccccccc|}
\hline
 & $\gamma$ & $\sigma$ & $\log _{10}\left(\lum_{\star} / \lum_{\odot}\right)$ & $\log _{10}\left(\phi_{\star} / (\mathrm{Mpc}^{-3} \rm{dex^{-1}}) \right)$ & $k_{L}$ & $k_{R 1}$ & $k_{R 2}$ \\
\hline
spiral & 1.0 & 0.50 & 9.78 & $-2.12$ & 4.49 & -0.54 & $-7.13$ \\
starburst & 1.0 & 0.35 & 11.17 & $-4.46$ & 1.96 & 3.79 & $-1.06$ \\
SF-AGN & 1.2 & 0.40 & 10.80 & $-3.20$ & 3.17 & 0.67 & $-3.17$ \\
\hline
\end{tabular}
\caption{Parameters entering the infrared luminosity function for star forming galaxies, for the three galaxy populations under consideration: spiral, starbursts, star-forming galaxies hosting an AGN.}
\label{tab:params_agn}
\end{table*}

The scaling relation between the GLF for energies between 0.1 $\mathrm{GeV}$ and $100 \ \mathrm{GeV}$, and the IR luminosity for wavelengths bewteen $8 \mu \mathrm{m}$ and $1000 \mu \mathrm{m}$ is given in Ref. \citep{Ackermann_2012_SFG} as:

\begin{equation}
\log _{10}\left(\frac{\lum_{0.1-100 \mathrm{GeV}}}{\mathrm{erg} \mathrm{s}^{-1}}\right)=\alpha_{\mathrm{IR}} \log _{10}\left(\frac{\lum_{8-1000 \mu \mathrm{m}}}{10^{10} \lum_{\odot}}\right)+\beta_{\mathrm{IR}}
\label{eq:L_to_L_ir}
\end{equation}
where $\alpha_{\mathrm{IR}}=1.09$ and $\beta_{\mathrm{IR}}=39.19$.
The GLF can thus be derived using Eqs. (\ref{eq:phi_hat_ir_SFG}) and (\ref{eq:L_to_L_ir}):
\begin{equation}
\phi_{\gamma}\left(\lum_{\gamma}, z\right)=\frac{\hat{\phi}_{\mathrm{IR}}}{\lum \  \log (10) } \frac{\mathrm{d} \log _{10}\left(\lum_{\mathrm{IR}}\right)}{\mathrm{d} \log_{10}\left(\lum_{\gamma}\right)}
\label{eq:phi_gamma_z_SFG}
\end{equation}

\section{Galaxies}
\label{app:galaxies}

For discrete tracers like galaxies, we need to model the way they populate each halo: the standard assumption is to separate the galaxies into central galaxies and satellite galaxies, thus defining a Halo Occupation Distribution (HOD) model \cite{Berlind:2001xk,Zheng:2004id}. The average number density of galaxies $\langle n_g(r | M,z) \rangle$ at radial position $r$, in a halo of mass $M$ and at redshift $z$, can be thus expressed as follows:
\begin{equation}
       \langle n_g(r | M,z) \rangle = \langle N_{\rm cen}(M,z) \rangle f_{\rm cen}(z) \,\delta_D(r)+ \langle N_{\rm sat}(M,z) \rangle \ u_{\rm sat}(r|M,z).
    \label{eq:HOD}
\end{equation}
where $\delta_D(r)$ is a Dirac delta function.
The HOD requires a survey-specific model for the average number of central galaxies $\langle {N}_{\rm cen}(M,z) \rangle$, the fraction of central galaxies $f_{\rm cen}(z)$, the average number of satellite galaxies $\langle{N}_{\rm sat}(M,z)\rangle$ and the density profile of the satellite halos hosting such galaxies $u_{\rm sat}(r|M,z)$. 

In our analysis, we consider a low-redshift galaxy catalog modeled on the characteristics of 2MRS \citep{Huchra_2MRS_2012,Ando_2MRS_2014,Ando_2MRS_2018}. We assume $f_{\rm cen}(z) = 1$, meaning that each halo hosts on average at most one central galaxy, independently of redshift. $u_{\rm sat}(r|M,z)$ is modeled according to the NFW density profile, as done for the halo dark matter profile discussed above in Sec. \ref{app:dark_matter}, but with a concentration parameter $c_{\rm{sat}}$ which is rescaled from the halo one as: $c_{\rm sat}(M,z) = b_{m}/b_g \cdot c_{\rm vir}(M,z)$, with $b_{m} = 6.7$ e $b_{g} = 0.66$ \citep{Ando_2MRS_2018}. For the number of central and satellite galaxies for 2MRS we adopt \citep{Ando_2MRS_2018}:
\begin{eqnarray}
 \langle N_{\rm cen}(M,z)\rangle &=& \frac{1}{2}
  \left[1+\mbox{erf}\left(\frac{\log M(z) - \log M_{\rm min}}{\sigma_{\log
		     M}}\right)\right]\,,\\
  \langle N_{\rm sat}(M,z)\rangle &=& \left(\frac{M(z)-M_{0}}{M_1}\right)^\alpha
  \Theta(M(z)-M_{0})\,, \label{eq:HOD_2mrs}
\end{eqnarray}
where masses are intended in units of solar masses and: $\log_{10} M_{\rm min} = 11.85$, $\log_{10} M_{\rm 1} = 11.97$, $\sigma_{\log M} = 0.15$, $\alpha = 0.846$ and we have set $M_0 = M_{\rm min}$.

Finally, the remaining quantities that define our modeling for the 2MRS galaxy catalog are its redshift interval, which we take as $z_{\rm min} = 0.0012$ and $z_{\rm max} = 0.1$, and the average number density of galaxies per unit volume $\bar{n}_{\rm{g}}(z)$, related to the average number density of galaxies per unit redshift $dN_g/dz$ as:
\begin{equation}
    \bar{n}_{\rm{g}}(z) = \frac{dN_g}{dV} = \frac{dN_g}{dz} \left[\frac{4 \pi c}{H(z)} \frac{ D^2_A(z)}{(1+z)^2}\right]^{-1} 
\end{equation}
where $D_A(z)$ is the angular diameter distance and $H(z)$ the Hubble rate.  For $\bar{n}_{\rm{g}}(z)$ we adopt the parameterization of Ref. \citep{Ando_2MRS_2018}, where: 

\begin{equation}
    \frac{d N_{\mathrm{g}}}{d z}=\frac{N_{\mathrm{g}} \beta}{z_{0}\, \Gamma[(m+1) / \beta]}\left(\frac{z}{z_{0}}\right)^{m} \exp \left[-\left(\frac{z}{z_{0}}\right)^{\beta}\right]
\end{equation}
with $N_g = 43182$ the total number of galaxies of the survey, $\beta = 1.64$, $z_0 =  0.0266$, $m= 1.31$ and  $\Gamma$ the Gamma function.
Using these quantities, one has an average angular number density for the survey equal to 3920 galaxies/sr$^{-2}$.

\section{3D power spectra}
\label{app:3D_PS}

The 1-halo and 2-halo terms for the 3D power spectra depend on which fields are being correlated. For the galaxies auto-correlation, we have  \cite{fornengo2014}:
\begin{align*}
     P^{\rm g\rm g}_{1h}(k,z) &= \int_{M_{\rm min}}^{M_{\rm max}} \de M \ \frac{dn}{dM}\frac{\langle N_{\rm g}\,(N_{\rm g}-1)\rangle}{\bar n_{\rm g}^2} \tilde v(k|M)^2 \\
 P^{\rm g\rm g}_{2h}(k,z) &= \left[\int_{m_{\rm min}}^{m_{\rm max}} \de M \,\frac{dn}{dM} b_h(M,z) \frac{\langle N_{\rm g}\rangle}{\bar n_{\rm g}}  \tilde v(k|M)  \right]^2\,P^{\rm lin}(k,z)
\end{align*}
where $\tilde v(k|M)$ is the Fourier transform of $\rho({\bm r}|M)/\bar{\rho}$ with $\bar{\rho}$ the current mean value of the matter density of the Universe, including both baryons and CDM, $b_h(M,z)$ is the halo bias, modeled as in Ref. \citep{sheth_tormen_2001}, $N_g$ the number of galaxies of the HOD, 
$\bar n_g = \int \de M\; \de n/\de M \;\langle N_g\rangle$ and $P^{\rm lin}(k)$ is the matter linear power spectrum, which we have calculated by using the package \texttt{pyccl} \citep{pyccl_2019}, based on \texttt{CAMB} \citep{camb_2011}.

When we cross-correlated galaxies with the gamma-ray emission from DM annihilation (which depends on the square of the dark matter density), the power spectra are \cite{fornengo2014}:
\begin{align*}
P^{\rm g\delta^2}_{1h}(k,z) &= \int_{M_{\rm min}}^{M_{\rm max}} \de M \ \frac{dn}{dM} \frac{\langle N_{\rm g}\rangle}{\bar n_{\rm g}} \tilde v(k|M) \frac{\tilde u(k|M)}{\Delta^2} \\
P^{\rm g\delta^2}_{2h}(k,z) &= \left[\int_{M_{\rm min}}^{M_{\rm max}} \de M \,\frac{dn}{dM} b_h(M,z) \frac{\langle N_{\rm g}\rangle}{\bar n_{\rm g}} \tilde v(k|M)]\right]\times  \\ &\times \left[\int_{M_{\rm min}}^{M_{\rm max}} \de M \,\frac{dn}{dM} b_h(M,z) \frac{\tilde u(k|M)}{\Delta^2}  \right]\,P^{\rm lin}(k,z)
\end{align*}
where $\tilde u(k|M)$ is the Fourier transform of {\rm $\rho^2({\bm r}|M)/\langle \rho(z) \rangle^2$} and $\Delta(z)$ the clumping factor introduced in Sec. \ref{app:dark_matter}.

The autocorrelation of gamma-ray emission from DM annihilation has \cite{fornengo2014}:
\begin{align*}
    & P^{\delta^2\delta^2}_{1h}(k,z) = \int_{M_{\rm min}} ^{M_{\rm max}} dM \ \frac{dn}{dM} \left( \frac{\tilde u(k|M)}{\Delta^2} \right)^2
    \\
    & P^{\delta^2\delta^2}_{2h}(k,z) =  
    \left[\int_{M_{\rm min}} ^{M_{\rm max}} dM \ \frac{dn}{dM} b_h(M,z) \frac{\tilde u(k|M)}{\Delta^2}\right]^2 \; P_{\rm lin}(k,z)
\end{align*}

Considering instead gamma-ray emission from astrophysical sources, we have several terms. The cross-correlation with galaxies reads \cite{fornengo2014}:
\begin{align*}
P^{Sg}_{1h}(k,z) &= 
\int_{\mathcal{L}_{\rm min}}^{\mathcal{L}_{\rm max}(z)}d\mathcal{L} \, 
\phi(\mathcal{L},z) \frac{\mathcal{L}}{\langle g_{S} \rangle} \ \frac{\langle N_g \rangle}{\bar{n}_g} 
\tilde v(k|M)
\\
P^{S\rm g}_{2h}(k,z) &= 
\left[\int_{\mathcal{L}_{\rm min}}^{\mathcal{L}_{\rm max}(z)} d \mathcal{L} \, 
\phi(\mathcal{L},z)\, b_{S}(\mathcal{L},z) 
\frac{\mathcal{L}}{\langle g_{S} \rangle} \right] \times \\
&\times \left[ \int_{M_{\rm min}}^{M_{\rm max}} d M \,\frac{d n}{d M} b_h(M,z) \frac{\langle N_g \rangle}{\bar{n}_g}
\tilde v(k|M)\right] P_{\rm lin}(k,z).
\end{align*}
where we have labeled the astrophysical sources with $S$. $\cal L$ denotes the luminosity of the source, $\phi(\mathcal{L},z)$ is the gamma-ray luminosity function, $\langle g_S \rangle = \int_{\mathcal{L}_{\rm min}}^{\mathcal{L}_{\rm max}(z)} d \mathcal{L}\;\phi(\mathcal{L},z) \;{\cal L}$ and finally $b_S(M,z)$ the source bias. The source bias is obtained from $b_h(M,z)$ \citep{sheth_tormen_2001}, with the mass-luminosity conversions $M(\mathcal{L})$ discussed in Sec. \ref{sec:Mass_luminosity_relation}. i.e. $b_S(M,z) = b_h(M(\mathcal{L}),z)$.

$\mathcal{L}_{\rm min}$ and $\mathcal{L}_{\rm max}(z)$, as well all other quantities referring to the astrophysical sources, are defined and introduced for each class of sources in Sec. \ref{app:astro_sources}.
Then, the correlation between the gamma-ray emission of astrophysical sources and the gamma-ray emission from dark matter annihilation reads \cite{fornengo2014}:
\begin{align*}
P^{S\delta^2}_{1h}(k,z) &= 
\int_{\mathcal{L}_{\rm min}}^{\mathcal{L}_{\rm max}(z)}d\mathcal{L} \, 
\phi(\mathcal{L},z) \frac{\mathcal{L}}{\langle g_{S} \rangle}
\frac{\tilde u(k|M)}{\Delta^2}
\\
P^{S\delta^2}_{2h}(k,z) &= 
\left[\int_{\mathcal{L}_{\rm min}}^{\mathcal{L}_{\rm max}(z)}d\mathcal{L} \, 
\phi(\mathcal{L},z)b_{S}(\mathcal{L},z)
\frac{\mathcal{L}}{\langle g_{S} \rangle} \right] \times \\
&\times \left[ \int_{M_{\rm min}}^{M_{\rm max}}d M \, \frac{d n}{d M} b_h(M,z) 
\frac{\tilde u(k|M)}{\Delta^2}\right]P_{\rm lin}(k,z).
\nonumber
\end{align*}
Finally, the correlation between the gamma-ray emission of astrophysical sources among themselves is \cite{camera2015tomographic,fornengo2014}:
\begin{align}
P^{S_i S_j}_{1h}(k,z)&= \int_{\mathcal{L}_{\rm min}}^{\mathcal{L}_{\rm max}(z)}
d \mathcal{L}\,\phi_i(\mathcal{L},z)
\left( \frac{\mathcal{L}}{\langle g_{S,i} \rangle} \right)^2\,\delta_{ij}
\\
P^{S_i S_j}_{2h}(k,z)&= \left[\int_{\mathcal{L}_{\rm min}}^{\mathcal{L}_{\rm max}(z)}\!\!
d\mathcal{L} \, \phi_i(\mathcal{L},z) \, b_{S,i}(\mathcal{L},z)\frac{\mathcal{L}}{\langle g_{S,i} \rangle} \right] \left[\int_{\mathcal{L}_{\rm min}}^{\mathcal{L}_{\rm max}(z)}\!\!
d\mathcal{L} \, \phi_j(\mathcal{L},z) \, b_{S,j}(\mathcal{L},z)\frac{\mathcal{L}}{\langle g_{S,j} \rangle} \right] P_{\rm lin}(k,z).
\label{eq:power spectrumB}
\end{align}
where $i,j$ refer to BLA, mAGN, FSRQ and SFG.

\section{\Fermi\  finite resolution}
\label{app:Fermi_PSF}

The finite angular resolution of the \Fermi\  affects small angular scales and its effect is obtained through a convolution of the predicted incoming gamma-ray flux with a opening-angle dependent point-spread function (PSF), \citep{ackermann2018} 
which depends on energy. In harmonic space, the convolution transforms into a product of the harmonic power spectrum of the gamma-ray signal with the spherical harmonic transform of the PSF, for which we adopt the parameterization derived in Ref. \citep{pinetti2019}, based on the measured \Fermi\ PSF \citep{ackermann2018}
\begin{equation}
     B_{\ell}(E_b) = \exp\left(- \frac{\sigma_b\left(\ell, E_b\right)^2 \ell^2}{2}\right).\label{eq:beam1}
\end{equation}
where :
\begin{equation}
\sigma_{b}(\ell, E_b) = \sigma_0^{\rm Fermi}(E_b) \left[1 + 0.25\,\sigma_0^{\rm Fermi}(E_b)\,\ell\right]^{-1}.
\label{eq:beam2}
\end{equation}
with
\begin{equation}
\sigma_0^{\rm Fermi}(E_b) = \left(\sigma_0^{\rm Fermi}(E_{\rm ref})  \times (E_b/E_{\rm ref})^{-0.95} + 0.05\right) \frac{\pi}{180} \left[{\rm rad}\right].
\label{eq:beam3}
\end{equation}
where we have defined $E_b = \sqrt{E_{{\rm min},b} \, E_{{\rm max},b}}$ as the reference energy of the $b^{\rm th}$ energy bin, being $E_{{\rm min},b}$ and $E_{{\rm max},b}$ its boundary energies. The energy bins adopted in our analysis and the values of $\sigma_0^{\rm Fermi}(E_{\rm ref})$ are listed in Table \ref{tab:Fermi_energy_bins}.

\acknowledgments
NF gratefully acknowledges for the hospitality of the Flatiron Institute Center for Computational Astrophysics (CCA) of the Simons Foundation and of the New York University, where part of this work has been completed.
We gratefully acknowledge D. Alonso, S. Arcari, F. Gnesutta, E. Pinetti, M. Regis and F. Urban and for useful exchanges and discussions. 
%
AR acknowledges that this paper and related research have been conducted at the operational site of the University of Turin - Department of Physics,  while attending the PhD programme in Space Science and Technology at the University of Trento (\href{https://phd.unitn.it/phd-sst/}{https://phd.unitn.it/phd-sst/}), Cycle XXXVIII, with the support of a scholarship financed by the Ministerial Decree no. 351 of 9th April 2022, based on the NRRP - funded by the European Union - NextGenerationEU - Mission 4 \enquote{Education and Research}, Component 1 \enquote{Enhancement of the offer of educational services: from nurseries to universities} - Investment 4.1 \enquote{Extension of the number of research doctorates and innovative doctorates for public administration and cultural heritage} [CUP E63C22001340001]. AR and NF acknowledge support from the Research grant TAsP (Theoretical Astroparticle Physics) funded by INFN. SC acknowledges support from the Italian Ministry of University and Research (MUR), PRIN 2022 `EXSKALIBUR – Euclid-Cross-SKA: Likelihood Inference Building for Universe's Research', Grant No.\ 20222BBYB9, CUP D53D2300252 0006, and from the European Union -- Next Generation EU. NF acknowledges support from the Italian Ministry of University and Research (MUR) via the PRIN 2022 Project No. 20228WHTYC, CUP C53C24000760006.
We finally acknowledge the support of the Computational Infrastructure for Science {\it Pleiadi} \cite{Pleiadi1, Pleiadi2} of the Italian National Institute for Astrophysics (INAF).

 ------------------------------

\bibliographystyle{JHEP}
\bibliography{biblioref}

\providecommand{\href}[2]{#2}\begingroup\raggedright\begin{thebibliography}{10}

\bibitem{Cirelli:2024ssz}
M.~Cirelli, A.~Strumia and J.~Zupan, \emph{{Dark Matter}},  \href{https://arxiv.org/abs/2406.01705}{{\ttfamily 2406.01705}}.

\bibitem{camera2013novel}
S.~Camera, M.~Fornasa, N.~Fornengo and M.~Regis, \emph{A novel approach in the weakly interacting massive particle quest: Cross-correlation of gamma-ray anisotropies and cosmic shear}, {\emph{The Astrophysical Journal Letters} {\bfseries 771} (2013) L5}.

\bibitem{camera2015tomographic}
S.~Camera, M.~Fornasa, N.~Fornengo and M.~Regis, \emph{Tomographic-spectral approach for dark matter detection in the cross-correlation between cosmic shear and diffuse $\gamma$-ray emission}, {\emph{Journal of Cosmology and Astroparticle Physics} {\bfseries 2015} (2015) 029}.

\bibitem{fornengo2014}
N.~Fornengo and M.~Regis, \emph{{Particle dark matter searches in the anisotropic sky}}, \href{https://doi.org/10.3389/fphy.2014.00006}{\emph{Front. Physics} {\bfseries 2} (2014) 6} [\href{https://arxiv.org/abs/1312.4835}{{\ttfamily 1312.4835}}].

\bibitem{fornengo2015evidence}
N.~Fornengo, L.~Perotto, M.~Regis and S.~Camera, \emph{Evidence of cross-correlation between the cmb lensing and the $\gamma$-ray sky}, {\emph{The Astrophysical journal letters} {\bfseries 802} (2015) L1}.

\bibitem{pinetti2019}
E.~Pinetti, S.~Camera, N.~Fornengo and M.~Regis, \emph{{Synergies across the spectrum for particle dark matter indirect detection: how HI intensity mapping meets gamma rays}}, \href{https://doi.org/10.1088/1475-7516/2020/07/044}{\emph{JCAP} {\bfseries 07} (2020) 044} [\href{https://arxiv.org/abs/1911.04989}{{\ttfamily 1911.04989}}].

\bibitem{Arcari:2022zul}
S.~Arcari, E.~Pinetti and N.~Fornengo, \emph{{Got plenty of nothing: cosmic voids as a probe of particle dark matter}}, \href{https://doi.org/10.1088/1475-7516/2022/11/011}{\emph{JCAP} {\bfseries 11} (2022) 011} [\href{https://arxiv.org/abs/2205.03360}{{\ttfamily 2205.03360}}].

\bibitem{shirasaki2014cross}
M.~Shirasaki, S.~Horiuchi and N.~Yoshida, \emph{Cross correlation of cosmic shear and extragalactic gamma-ray background: Constraints on the dark matter annihilation cross section}, {\emph{Physical Review D} {\bfseries 90} (2014) 063502}.

\bibitem{shirasaki2016cosmological}
M.~Shirasaki, O.~Macias, S.~Horiuchi, S.~Shirai and N.~Yoshida, \emph{Cosmological constraints on dark matter annihilation and decay: Cross-correlation analysis of the extragalactic $\gamma$-ray background and cosmic shear}, {\emph{Physical Review D} {\bfseries 94} (2016) 063522}.

\bibitem{troster2017cross}
T.~Tr{\"o}ster, S.~Camera, M.~Fornasa, M.~Regis, L.~Van~Waerbeke, J.~Harnois-D{\'e}raps et~al., \emph{Cross-correlation of weak lensing and gamma rays: implications for the nature of dark matter}, {\emph{Monthly Notices of the Royal Astronomical Society} {\bfseries 467} (2017) 2706}.

\bibitem{ammazzalorso2020}
{\scshape DES} collaboration, \emph{{Detection of Cross-Correlation between Gravitational Lensing and $\gamma$ Rays}}, \href{https://doi.org/10.1103/PhysRevLett.124.101102}{\emph{Phys. Rev. Lett.} {\bfseries 124} (2020) 101102} [\href{https://arxiv.org/abs/1907.13484}{{\ttfamily 1907.13484}}].

\bibitem{Thakore_DES_2025}
B.~{Thakore}, M.~{Negro}, M.~{Regis}, S.~{Camera}, D.~{Gruen}, N.~{Fornengo} et~al., \emph{{High-significance detection of correlation between the unresolved gamma-ray background and the large-scale cosmic structure}}, \href{https://doi.org/10.1088/1475-7516/2025/06/037}{\emph{\jcap} {\bfseries 2025} (2025) 037} [\href{https://arxiv.org/abs/2501.10506}{{\ttfamily 2501.10506}}].

\bibitem{xia2011}
J.-Q.~Xia, A.~Cuoco, E.~Branchini, M.~Fornasa and M.~Viel, \emph{{A cross-correlation study of the Fermi-LAT $\gamma$-ray diffuse extragalactic signal}}, \href{https://doi.org/10.1111/j.1365-2966.2011.19200.x}{\emph{Mon. Not. Roy. Astron. Soc.} {\bfseries 416} (2011) 2247} [\href{https://arxiv.org/abs/1103.4861}{{\ttfamily 1103.4861}}].

\bibitem{xia2015tomography}
J.-Q.~Xia, A.~Cuoco, E.~Branchini and M.~Viel, \emph{Tomography of the fermi-lat $\gamma$-ray diffuse extragalactic signal via cross correlations with galaxy catalogs}, {\emph{The Astrophysical Journal Supplement Series} {\bfseries 217} (2015) 15}.

\bibitem{regis2015particle}
M.~Regis, J.-Q.~Xia, A.~Cuoco, E.~Branchini, N.~Fornengo and M.~Viel, \emph{Particle dark matter searches outside the local group}, {\emph{Physical Review Letters} {\bfseries 114} (2015) 241301}.

\bibitem{cuoco2015dark}
A.~Cuoco, J.-Q.~Xia, M.~Regis, E.~Branchini, N.~Fornengo and M.~Viel, \emph{Dark matter searches in the gamma-ray extragalactic background via cross-correlations with galaxy catalogs}, {\emph{The Astrophysical Journal Supplement Series} {\bfseries 221} (2015) 29}.

\bibitem{shirasaki2015cross}
M.~Shirasaki, S.~Horiuchi and N.~Yoshida, \emph{Cross-correlation of the extragalactic gamma-ray background with luminous red galaxies}, {\emph{Physical Review D} {\bfseries 92} (2015) 123540}.

\bibitem{cuoco2017tomographic}
A.~Cuoco, M.~Bilicki, J.-Q.~Xia and E.~Branchini, \emph{Tomographic imaging of the fermi-lat $\gamma$-ray sky through cross-correlations: A wider and deeper look}, {\emph{The Astrophysical Journal Supplement Series} {\bfseries 232} (2017) 10}.

\bibitem{Ammazzalorso:2018evf}
S.~Ammazzalorso, N.~Fornengo, S.~Horiuchi and M.~Regis, \emph{{Characterizing the local gamma-ray Universe via angular cross-correlations}}, \href{https://doi.org/10.1103/PhysRevD.98.103007}{\emph{Phys. Rev. D} {\bfseries 98} (2018) 103007} [\href{https://arxiv.org/abs/1808.09225}{{\ttfamily 1808.09225}}].

\bibitem{paopiamsap2024constraints}
A.~Paopiamsap, D.~Alonso, D.J.~Bartlett and M.~Bilicki, \emph{Constraints on dark matter and astrophysics from tomographic $\gamma$-ray cross-correlations}, {\emph{Physical Review D} {\bfseries 109} (2024) 103517}.

\bibitem{branchini2017cross}
E.~Branchini, S.~Camera, A.~Cuoco, N.~Fornengo, M.~Regis, M.~Viel et~al., \emph{Cross-correlating the $\gamma$-ray sky with catalogs of galaxy clusters}, {\emph{The Astrophysical Journal Supplement Series} {\bfseries 228} (2017) 8}.

\bibitem{shirasaki2018correlation}
M.~Shirasaki, O.~Macias, S.~Horiuchi, N.~Yoshida, C.-H.~Lee and A.J.~Nishizawa, \emph{Correlation of extragalactic $\gamma$ rays with cosmic matter density distributions from weak gravitational lensing}, {\emph{Physical Review D} {\bfseries 97} (2018) 123015}.

\bibitem{hashimoto2019measurement}
D.~Hashimoto, A.J.~Nishizawa, M.~Shirasaki, O.~Macias, S.~Horiuchi, H.~Tashiro et~al., \emph{Measurement of redshift-dependent cross-correlation of hsc clusters and fermi $\gamma$-rays}, {\emph{Monthly Notices of the Royal Astronomical Society} {\bfseries 484} (2019) 5256}.

\bibitem{colavincenzo2020searching}
M.~Colavincenzo, X.~Tan, S.~Ammazzalorso, S.~Camera, M.~Regis, J.-Q.~Xia et~al., \emph{Searching for gamma-ray emission from galaxy clusters at low redshift}, {\emph{Monthly Notices of the Royal Astronomical Society} {\bfseries 491} (2020) 3225}.

\bibitem{tan2020bounds}
X.~Tan, M.~Colavincenzo and S.~Ammazzalorso, \emph{Bounds on wimp dark matter from galaxy clusters at low redshift}, {\emph{Monthly Notices of the Royal Astronomical Society} {\bfseries 495} (2020) 114}.

\bibitem{Ando_2MRS_2014}
S.~{Ando}, A.~{Benoit-L{\'e}vy} and E.~{Komatsu}, \emph{{Mapping dark matter in the gamma-ray sky with galaxy catalogs}}, \href{https://doi.org/10.1103/PhysRevD.90.023514}{\emph{\prd} {\bfseries 90} (2014) 023514} [\href{https://arxiv.org/abs/1312.4403}{{\ttfamily 1312.4403}}].

\bibitem{fornasa2016angular}
M.~Fornasa, A.~Cuoco, J.~Zavala, J.M.~Gaskins, M.A.~S{\'a}nchez-Conde, G.~Gomez-Vargas et~al., \emph{Angular power spectrum of the diffuse gamma-ray emission as measured by the fermi large area telescope and constraints on its dark matter interpretation}, {\emph{Physical Review D} {\bfseries 94} (2016) 123005}.

\bibitem{feng2017planck}
C.~Feng, A.~Cooray and B.~Keating, \emph{Planck lensing and cosmic infrared background cross-correlation with fermi-lat: Tracing dark matter signals in the gamma-ray background}, {\emph{The Astrophysical Journal} {\bfseries 836} (2017) 127}.

\bibitem{Tan:2020fbc}
X.-H.~Tan, J.-P.~Dai and J.-Q.~Xia, \emph{{Searching for Integrated Sachs\textendash{}Wolfe Effect from $Fermi$-LAT diffuse $\gamma$-ray map}}, \href{https://doi.org/10.1016/j.dark.2020.100585}{\emph{Phys. Dark Univ.} {\bfseries 29} (2020) 100585} [\href{https://arxiv.org/abs/2005.03833}{{\ttfamily 2005.03833}}].

\bibitem{Zhou:2024cld}
B.~Zhou, J.L.~Bernal, E.~Pinetti, H.A.G.~Cruz and M.~Kamionkowski, \emph{{Cross Correlating the Unresolved Gamma-Ray Background with Cosmic Large-Scale Structure from DESI: Implications for Astrophysics and Dark Matter}},  \href{https://arxiv.org/abs/2410.00375}{{\ttfamily 2410.00375}}.

\bibitem{Pinetti:2025qpi}
E.~Pinetti, V.~Vodeb, A.~Amerio, A.~Cuoco, S.~Camera, N.~Fornengo et~al., \emph{{Across the Universe: Dark Matter and Galaxy Cross-Correlations with the Cherenkov Telescope Array Observatory}},  \href{https://arxiv.org/abs/2505.20383}{{\ttfamily 2505.20383}}.

\bibitem{Alonso_Wiener_GW}
D.~{Alonso}, G.~{Cusin}, P.G.~{Ferreira} and C.~{Pitrou}, \emph{{Detecting the anisotropic astrophysical gravitational wave background in the presence of shot noise through cross-correlations}}, \href{https://doi.org/10.1103/PhysRevD.102.023002}{\emph{\prd} {\bfseries 102} (2020) 023002} [\href{https://arxiv.org/abs/2002.02888}{{\ttfamily 2002.02888}}].

\bibitem{Urban_Alonso_Camera_Wiener}
F.R.~{Urban}, S.~{Camera} and D.~{Alonso}, \emph{{Detecting ultra-high-energy cosmic ray anisotropies through harmonic cross-correlations}}, \href{https://doi.org/10.1051/0004-6361/202038459}{\emph{\aap} {\bfseries 652} (2021) A41} [\href{https://arxiv.org/abs/2005.00244}{{\ttfamily 2005.00244}}].

\bibitem{Wiener_1949}
N.~Wiener, \emph{{Extrapolation, Interpolation, and Smoothing of Stationary Time Series: With Engineering Applications}}, The MIT Press (08, 1949), \href{https://doi.org/10.7551/mitpress/2946.001.0001}{10.7551/mitpress/2946.001.0001}.

\bibitem{Rybicky_Press_1992}
G.B.~{Rybicki} and W.H.~{Press}, \emph{{Interpolation, Realization, and Reconstruction of Noisy, Irregularly Sampled Data}}, \href{https://doi.org/10.1086/171845}{\emph{\apj} {\bfseries 398} (1992) 169}.

\bibitem{Huchra_2MRS_2012}
J.P.~{Huchra}, L.M.~{Macri}, K.L.~{Masters}, T.H.~{Jarrett}, P.~{Berlind}, M.~{Calkins} et~al., \emph{{The 2MASS Redshift Survey{\textemdash}Description and Data Release}}, \href{https://doi.org/10.1088/0067-0049/199/2/26}{\emph{\apjs} {\bfseries 199} (2012) 26} [\href{https://arxiv.org/abs/1108.0669}{{\ttfamily 1108.0669}}].

\bibitem{Planck_2018}
{Planck Collaboration}, N.~{Aghanim}, Y.~{Akrami}, M.~{Ashdown}, J.~{Aumont}, C.~{Baccigalupi} et~al., \emph{{Planck 2018 results. VI. Cosmological parameters}}, \href{https://doi.org/10.1051/0004-6361/201833910}{\emph{\aap} {\bfseries 641} (2020) A6} [\href{https://arxiv.org/abs/1807.06209}{{\ttfamily 1807.06209}}].

\bibitem{limber1953}
D.~Limber, \emph{{The analysis of counts of the extragalactic nebulae in terms of a fluctuating density field}}, \href{https://doi.org/10.1086/145672}{\emph{Astrophys. J.} {\bfseries 134} (1953) A6}.

\bibitem{limber1992}
N.~Kaiser, \emph{{Weak gravitational lensing of distant galaxies}}, \href{https://doi.org/10.1086/171151}{\emph{Astrophys. J.} {\bfseries 388} (1992) 272}.

\bibitem{limber1998}
N.~Kaiser, \emph{{Weak lensing and cosmology}}, \href{https://doi.org/10.1086/305515}{\emph{Astrophys. J.} {\bfseries 498} (1998) 26} [\href{https://arxiv.org/abs/astro-ph/9610120}{{\ttfamily astro-ph/9610120}}].

\bibitem{cooray2002}
A.~Cooray and R.K.~Sheth, \emph{{Halo Models of Large Scale Structure}}, \href{https://doi.org/10.1016/S0370-1573(02)00276-4}{\emph{Phys. Rept.} {\bfseries 372} (2002) 1} [\href{https://arxiv.org/abs/astro-ph/0206508}{{\ttfamily astro-ph/0206508}}].

\bibitem{Asgari_halo_model_2023}
M.~{Asgari}, A.J.~{Mead} and C.~{Heymans}, \emph{{The halo model for cosmology: a pedagogical review}}, \href{https://doi.org/10.21105/astro.2303.08752}{\emph{The Open Journal of Astrophysics} {\bfseries 6} (2023) 39} [\href{https://arxiv.org/abs/2303.08752}{{\ttfamily 2303.08752}}].

\bibitem{Franceschini_2008}
A.~{Franceschini}, G.~{Rodighiero} and M.~{Vaccari}, \emph{{Extragalactic optical-infrared background radiation, its time evolution and the cosmic photon-photon opacity}}, \href{https://doi.org/10.1051/0004-6361:200809691}{\emph{\aap} {\bfseries 487} (2008) 837} [\href{https://arxiv.org/abs/0805.1841}{{\ttfamily 0805.1841}}].

\bibitem{Arina_Cosmix_2023}
C.~{Arina}, M.~{Di Mauro}, N.~{Fornengo}, J.~{Heisig}, A.~{Jueid} and R.~{Ruiz de Austri}, \emph{{CosmiXs: cosmic messenger spectra for indirect dark matter searches}}, \href{https://doi.org/10.1088/1475-7516/2024/03/035}{\emph{\jcap} {\bfseries 2024} (2024) 035} [\href{https://arxiv.org/abs/2312.01153}{{\ttfamily 2312.01153}}].

\bibitem{ackermann2018}
{\scshape Fermi-LAT} collaboration, \emph{{Unresolved Gamma-Ray Sky through its Angular Power Spectrum}}, \href{https://doi.org/10.1103/PhysRevLett.121.241101}{\emph{Phys. Rev. Lett.} {\bfseries 121} (2018) 241101} [\href{https://arxiv.org/abs/1812.02079}{{\ttfamily 1812.02079}}].

\bibitem{Ando_2MRS_2018}
S.~{Ando}, A.~{Benoit-L{\'e}vy} and E.~{Komatsu}, \emph{{Angular power spectrum of galaxies in the 2MASS Redshift Survey}}, \href{https://doi.org/10.1093/mnras/stx2634}{\emph{\mnras} {\bfseries 473} (2018) 4318} [\href{https://arxiv.org/abs/1706.05422}{{\ttfamily 1706.05422}}].

\bibitem{2004MNRAS.347..645P}
W.J.~{Percival}, L.~{Verde} and J.A.~{Peacock}, \emph{{Fourier analysis of luminosity-dependent galaxy clustering}}, \href{https://doi.org/10.1111/j.1365-2966.2004.07245.x}{\emph{\mnras} {\bfseries 347} (2004) 645} [\href{https://arxiv.org/abs/astro-ph/0306511}{{\ttfamily astro-ph/0306511}}].

\bibitem{Multitracer_McDonald_Seljak_2009}
P.~McDonald and U.~Seljak, \emph{How to evade the sample variance limit on measurements of redshift-space distortions}, \href{https://doi.org/10.1088/1475-7516/2009/10/007}{\emph{Journal of Cosmology and Astroparticle Physics} {\bfseries 2009} (2009) 007}.

\bibitem{Multitracer_Seljak_2009}
U.~{Seljak}, \emph{{Extracting Primordial Non-Gaussianity without Cosmic Variance}}, \href{https://doi.org/10.1103/PhysRevLett.102.021302}{\emph{\prl} {\bfseries 102} (2009) 021302} [\href{https://arxiv.org/abs/0807.1770}{{\ttfamily 0807.1770}}].

\bibitem{Abramo_Leonard_2013_Multitracer}
L.R.~{Abramo} and K.E.~{Leonard}, \emph{{Why multitracer surveys beat cosmic variance}}, \href{https://doi.org/10.1093/mnras/stt465}{\emph{\mnras} {\bfseries 432} (2013) 318} [\href{https://arxiv.org/abs/1302.5444}{{\ttfamily 1302.5444}}].

\bibitem{2014MNRAS.442.2511F}
L.D.~{Ferramacho}, M.G.~{Santos}, M.J.~{Jarvis} and S.~{Camera}, \emph{{Radio galaxy populations and the multitracer technique: pushing the limits on primordial non-Gaussianity}}, \href{https://doi.org/10.1093/mnras/stu1015}{\emph{\mnras} {\bfseries 442} (2014) 2511} [\href{https://arxiv.org/abs/1402.2290}{{\ttfamily 1402.2290}}].

\bibitem{2015ApJ...812L..22F}
J.~{Fonseca}, S.~{Camera}, M.G.~{Santos} and R.~{Maartens}, \emph{{Hunting Down Horizon-scale Effects with Multi-wavelength Surveys}}, \href{https://doi.org/10.1088/2041-8205/812/2/L22}{\emph{\apjl} {\bfseries 812} (2015) L22} [\href{https://arxiv.org/abs/1507.04605}{{\ttfamily 1507.04605}}].

\bibitem{Hess:2021cdp}
{\scshape Hess, HAWC, VERITAS, MAGIC, H.E.S.S., Fermi-LAT} collaboration, \emph{{Combined dark matter searches towards dwarf spheroidal galaxies with Fermi-LAT, HAWC, H.E.S.S., MAGIC, and VERITAS}}, \href{https://doi.org/10.22323/1.395.0528}{\emph{PoS} {\bfseries ICRC2021} (2021) 528} [\href{https://arxiv.org/abs/2108.13646}{{\ttfamily 2108.13646}}].

\bibitem{Cross_Alimi_Valls_Blanchard}
J.M.~{Alimi}, D.~{Valls-Gabaud} and A.~{Blanchard}, \emph{{A cross-correlation analysis of luminosity segregation in the clustering of galaxies}}, {\emph{\aap} {\bfseries 206} (1988) L11}.

\bibitem{Theroleofcross_Blanchard}
A.~{Blanchard}, \emph{{The Role of Cross-Correlations in the Multi-Tracer Area}}, \href{https://doi.org/10.3390/universe8090479}{\emph{Universe} {\bfseries 8} (2022) 479}.

\bibitem{Fermi-LAT:2018udj}
{\scshape Fermi-LAT} collaboration, \emph{{Unresolved Gamma-Ray Sky through its Angular Power Spectrum}}, \href{https://doi.org/10.1103/PhysRevLett.121.241101}{\emph{Phys. Rev. Lett.} {\bfseries 121} (2018) 241101} [\href{https://arxiv.org/abs/1812.02079}{{\ttfamily 1812.02079}}].

\bibitem{Fisher_Lahav_Wiener}
K.B.~{Fisher}, O.~{Lahav}, Y.~{Hoffman}, D.~{Lynden-Bell} and S.~{Zaroubi}, \emph{{Wiener reconstruction of density, velocity and potential fields from all-sky galaxy redshift surveys}}, \href{https://doi.org/10.1093/mnras/272.4.885}{\emph{\mnras} {\bfseries 272} (1995) 885} [\href{https://arxiv.org/abs/astro-ph/9406009}{{\ttfamily astro-ph/9406009}}].

\bibitem{Lahav_Fisher_Wiener}
O.~{Lahav}, K.B.~{Fisher}, Y.~{Hoffman}, C.A.~{Scharf} and S.~{Zaroubi}, \emph{{Wiener Reconstruction of All-Sky Galaxy Surveys in Spherical Harmonics}}, \href{https://doi.org/10.1086/187244}{\emph{\apjl} {\bfseries 423} (1994) L93} [\href{https://arxiv.org/abs/astro-ph/9311059}{{\ttfamily astro-ph/9311059}}].

\bibitem{Zaroubi_Hoffman_Fisher_Wiener}
S.~{Zaroubi}, Y.~{Hoffman}, K.B.~{Fisher} and O.~{Lahav}, \emph{{Wiener Reconstruction of the Large-Scale Structure}}, \href{https://doi.org/10.1086/176070}{\emph{\apj} {\bfseries 449} (1995) 446} [\href{https://arxiv.org/abs/astro-ph/9410080}{{\ttfamily astro-ph/9410080}}].

\bibitem{}
A.~{Raichoor}, J.~{Moustakas}, J.A.~{Newman}, T.~{Karim}, S.~{Ahlen}, S.~{Alam} et~al., \emph{{Target Selection and Validation of DESI Emission Line Galaxies}}, \href{https://doi.org/10.3847/1538-3881/acb213}{\emph{\aj} {\bfseries 165} (2023) 126} [\href{https://arxiv.org/abs/2208.08513}{{\ttfamily 2208.08513}}].

\bibitem{urban2023modelling}
F.R.~Urban, D.~Alonso and S.~Camera, \emph{Modelling cross-correlations of ultra-high-energy cosmic rays and galaxies},  2023.

\bibitem{Bryan_Norman_1998}
G.L.~Bryan and M.L.~Norman, \emph{Statistical properties of x-ray clusters: Analytic and numerical comparisons}, \href{https://doi.org/10.1086/305262}{\emph{The Astrophysical Journal} {\bfseries 495} (1998) 80}.

\bibitem{navarro1996}
J.F.~Navarro, C.S.~Frenk and S.D.M.~White, \emph{{The Structure of cold dark matter halos}}, \href{https://doi.org/10.1086/177173}{\emph{Astrophys. J.} {\bfseries 462} (1996) 563} [\href{https://arxiv.org/abs/astro-ph/9508025}{{\ttfamily astro-ph/9508025}}].

\bibitem{NFW_1997}
J.F.~{Navarro}, C.S.~{Frenk} and S.D.M.~{White}, \emph{{A Universal Density Profile from Hierarchical Clustering}}, \href{https://doi.org/10.1086/304888}{\emph{\apj} {\bfseries 490} (1997) 493} [\href{https://arxiv.org/abs/astro-ph/9611107}{{\ttfamily astro-ph/9611107}}].

\bibitem{Correa_2015}
C.A.~Correa, J.S.B.~Wyithe, J.~Schaye and A.R.~Duffy, \emph{{The accretion history of dark matter haloes – III. A physical model for the concentration–mass relation}}, \href{https://doi.org/10.1093/mnras/stv1363}{\emph{Monthly Notices of the Royal Astronomical Society} {\bfseries 452} (2015) 1217} [\href{https://arxiv.org/abs/https://academic.oup.com/mnras/article-pdf/452/2/1217/18505290/stv1363.pdf}{{\ttfamily https://academic.oup.com/mnras/article-pdf/452/2/1217/18505290/stv1363.pdf}}].

\bibitem{Coe2010}
D.~{Coe}, \emph{{Dark Matter Halo Mass Profiles}}, \href{https://doi.org/10.48550/arXiv.1005.0411}{\emph{arXiv e-prints} (2010) } [\href{https://arxiv.org/abs/1005.0411}{{\ttfamily 1005.0411}}].

\bibitem{sheth_tormen_1999}
R.K.~Sheth and G.~Tormen, \emph{{Large scale bias and the peak background split}}, \href{https://doi.org/10.1046/j.1365-8711.1999.02692.x}{\emph{Mon. Not. Roy. Astron. Soc.} {\bfseries 308} (1999) 119} [\href{https://arxiv.org/abs/astro-ph/9901122}{{\ttfamily astro-ph/9901122}}].

\bibitem{Moline_Sanchez_Conde_2017}
{\'A}.~{Molin{\'e}}, M.A.~{S{\'a}nchez-Conde}, S.~{Palomares-Ruiz} and F.~{Prada}, \emph{{Characterization of subhalo structural properties and implications for dark matter annihilation signals}}, \href{https://doi.org/10.1093/mnras/stx026}{\emph{\mnras} {\bfseries 466} (2017) 4974} [\href{https://arxiv.org/abs/1603.04057}{{\ttfamily 1603.04057}}].

\bibitem{Hogg:1999ad}
D.W.~Hogg, \emph{{Distance measures in cosmology}},  \href{https://arxiv.org/abs/astro-ph/9905116}{{\ttfamily astro-ph/9905116}}.

\bibitem{Manconi:2019ynl}
S.~Manconi, M.~Korsmeier, F.~Donato, N.~Fornengo, M.~Regis and H.~Zechlin, \emph{{Testing gamma-ray models of blazars in the extragalactic sky}}, \href{https://doi.org/10.1103/PhysRevD.101.103026}{\emph{Phys. Rev. D} {\bfseries 101} (2020) 103026} [\href{https://arxiv.org/abs/1912.01622}{{\ttfamily 1912.01622}}].

\bibitem{Pinetti_thesis_2022}
E.~{Pinetti}, \emph{{From gamma rays to radio waves: Dark Matter searches across the spectrum}}, \href{https://doi.org/10.48550/arXiv.2212.00125}{\emph{arXiv e-prints} (2022) } [\href{https://arxiv.org/abs/2212.00125}{{\ttfamily 2212.00125}}].

\bibitem{ajello2014}
M.~Ajello et~al., \emph{{The Cosmic Evolution of Fermi BL Lacertae Objects}}, \href{https://doi.org/10.1088/0004-637X/780/1/73}{\emph{Astrophys. J.} {\bfseries 780} (2014) 73} [\href{https://arxiv.org/abs/1310.0006}{{\ttfamily 1310.0006}}].

\bibitem{ajello2012}
M.~Ajello et~al., \emph{{The Luminosity Function of Fermi-detected Flat-Spectrum Radio Quasars}}, \href{https://doi.org/10.1088/0004-637X/751/2/108}{\emph{Astrophys. J.} {\bfseries 751} (2012) 108} [\href{https://arxiv.org/abs/1110.3787}{{\ttfamily 1110.3787}}].

\bibitem{ackermann2015}
{\scshape Fermi-LAT} collaboration, \emph{{The spectrum of isotropic diffuse gamma-ray emission between 100 MeV and 820 GeV}}, \href{https://doi.org/10.1088/0004-637X/799/1/86}{\emph{Astrophys. J.} {\bfseries 799} (2015) 86} [\href{https://arxiv.org/abs/1410.3696}{{\ttfamily 1410.3696}}].

\bibitem{willot2001}
C.J.~Willott, S.~Rawlings, K.M.~Blundell, M.~Lacy and S.A.~Eales, \emph{{The radio luminosity function from the low-frequency 3crr, 6ce \& 7crs complete samples}}, \href{https://doi.org/10.1046/j.1365-8711.2001.04101.x}{\emph{Mon. Not. Roy. Astron. Soc.} {\bfseries 322} (2001) 536} [\href{https://arxiv.org/abs/astro-ph/0010419}{{\ttfamily astro-ph/0010419}}].

\bibitem{Inoue:2011bm}
Y.~Inoue, \emph{{Contribution of the Gamma-ray Loud Radio Galaxies Core Emissions to the Cosmic MeV and GeV Gamma-Ray Background Radiation}}, \href{https://doi.org/10.1088/0004-637X/733/1/66}{\emph{Astrophys. J.} {\bfseries 733} (2011) 66} [\href{https://arxiv.org/abs/1103.3946}{{\ttfamily 1103.3946}}].

\bibitem{Dimauro2014}
M.~Di~Mauro, F.~Calore, F.~Donato, M.~Ajello and L.~Latronico, \emph{{Diffuse $\gamma$-ray emission from misaligned active galactic nuclei}}, \href{https://doi.org/10.1088/0004-637X/780/2/161}{\emph{Astrophys. J.} {\bfseries 780} (2014) 161} [\href{https://arxiv.org/abs/1304.0908}{{\ttfamily 1304.0908}}].

\bibitem{Lara:2004ee}
L.~Lara, G.~Giovannini, W.D.~Cotton, L.~Feretti, J.M.~Marcaide, I.~Marquez et~al., \emph{{A New sample of large angular size radio galaxies. 3. Statistics and evolution of the grown population}}, \href{https://doi.org/10.1051/0004-6361:20035676}{\emph{Astron. Astrophys.} {\bfseries 421} (2004) 899} [\href{https://arxiv.org/abs/astro-ph/0404373}{{\ttfamily astro-ph/0404373}}].

\bibitem{gruppioni2013}
C.~Gruppioni et~al., \emph{{The Herschel PEP/HerMES Luminosity Function. I: Probing the Evolution of PACS selected Galaxies to z\textasciitilde{}4}}, \href{https://doi.org/10.1093/mnras/stt308}{\emph{Mon. Not. Roy. Astron. Soc.} {\bfseries 432} (2013) 23} [\href{https://arxiv.org/abs/1302.5209}{{\ttfamily 1302.5209}}].

\bibitem{Ackermann_2012_SFG}
M.~Ackermann, M.~Ajello, A.~Allafort, L.~Baldini, J.~Ballet, D.~Bastieri et~al., \emph{Gev observations of star-forming galaxies with the fermi large area telescope}, \href{https://doi.org/10.1088/0004-637X/755/2/164}{\emph{The Astrophysical Journal} {\bfseries 755} (2012) 164}.

\bibitem{Berlind:2001xk}
A.A.~Berlind and D.H.~Weinberg, \emph{{The Halo occupation distribution: Towards an empirical determination of the relation between galaxies and mass}}, \href{https://doi.org/10.1086/341469}{\emph{Astrophys. J.} {\bfseries 575} (2002) 587} [\href{https://arxiv.org/abs/astro-ph/0109001}{{\ttfamily astro-ph/0109001}}].

\bibitem{Zheng:2004id}
Z.~Zheng, A.A.~Berlind, D.H.~Weinberg, A.J.~Benson, C.M.~Baugh, S.~Cole et~al., \emph{{Theoretical models of the halo occupation distribution: Separating central and satellite galaxies}}, \href{https://doi.org/10.1086/466510}{\emph{Astrophys. J.} {\bfseries 633} (2005) 791} [\href{https://arxiv.org/abs/astro-ph/0408564}{{\ttfamily astro-ph/0408564}}].

\bibitem{sheth_tormen_2001}
R.K.~Sheth, H.J.~Mo and G.~Tormen, \emph{{Ellipsoidal collapse and an improved model for the number and spatial distribution of dark matter haloes}}, \href{https://doi.org/10.1046/j.1365-8711.2001.04006.x}{\emph{Mon. Not. Roy. Astron. Soc.} {\bfseries 323} (2001) 1} [\href{https://arxiv.org/abs/astro-ph/9907024}{{\ttfamily astro-ph/9907024}}].

\bibitem{pyccl_2019}
N.E.~{Chisari}, D.~{Alonso}, E.~{Krause}, C.D.~{Leonard}, P.~{Bull}, J.~{Neveu} et~al., \emph{{Core Cosmology Library: Precision Cosmological Predictions for LSST}}, \href{https://doi.org/10.3847/1538-4365/ab1658}{\emph{\apjs} {\bfseries 242} (2019) 2} [\href{https://arxiv.org/abs/1812.05995}{{\ttfamily 1812.05995}}].

\bibitem{camb_2011}
A.~{Lewis} and A.~{Challinor}, ``{CAMB: Code for Anisotropies in the Microwave Background}.'' Astrophysics Source Code Library, record ascl:1102.026, Feb., 2011.

\bibitem{Pleiadi1}
S.~{Bertocco}, D.~{Goz}, L.~{Tornatore}, A.~{Ragagnin}, G.~{Maggio}, F.~{Gasparo} et~al., \emph{{INAF Trieste Astronomical Observatory Information Technology Framework}},  in \emph{Astronomical Data Analysis Software and Systems XXIX}, R.~{Pizzo}, E.R.~{Deul}, J.D.~{Mol}, J.~{de Plaa} and H.~{Verkouter}, eds., vol.~527 of \emph{Astronomical Society of the Pacific Conference Series}, p.~303, Jan., 2020, \href{https://doi.org/10.48550/arXiv.1912.05340}{DOI} [\href{https://arxiv.org/abs/1912.05340}{{\ttfamily 1912.05340}}].

\bibitem{Pleiadi2}
G.~{Taffoni}, U.~{Becciani}, B.~{Garilli}, G.~{Maggio}, F.~{Pasian}, G.~{Umana} et~al., \emph{{CHIPP: INAF Pilot Project for HTC, HPC and HPDA}},  in \emph{Astronomical Data Analysis Software and Systems XXIX}, R.~{Pizzo}, E.R.~{Deul}, J.D.~{Mol}, J.~{de Plaa} and H.~{Verkouter}, eds., vol.~527 of \emph{Astronomical Society of the Pacific Conference Series}, p.~307, Jan., 2020, \href{https://doi.org/10.48550/arXiv.2002.01283}{DOI} [\href{https://arxiv.org/abs/2002.01283}{{\ttfamily 2002.01283}}].

\end{thebibliography}\endgroup

\end{document}